\documentclass[pre,showpacs,amsmath,twocolumn]{revtex4}

\usepackage{graphicx}
\usepackage{epsfig}
\include{graphics}

\begin{document}
\title{Strong Collapse Turbulence  in Quintic Nonlinear Schr\"odinger Equation}

\author{Yeojin Chung$^{1}$ and Pavel M. Lushnikov$^{2}$}

\affiliation{$^1$Department of Mathematics, Southern Methodist University, Dallas, Texas, USA \\
  $^2$ Department of Mathematics and Statistics, University of New Mexico, Albuquerque, New Mexico, USA
}


\date{
\today}

\begin{abstract}
We consider the quintic one dimensional nonlinear Schr\"odinger equation with forcing and both linear and nonlinear dissipation.
Quintic nonlinearity results in multiple collapse events randomly distributed in space and time forming forced turbulence. Without dissipation each of these collapses produces finite time singularity but
dissipative terms prevents actual formation of singularity. In statistical steady state of the developed turbulence the spatial correlation function has a universal form with the correlation length determined by the
modulational instability scale.  The amplitude fluctuations at that scale are nearly-Gaussian while the large amplitude tail of probability density function (PDF) is strongly non-Gaussian with power-like behavior.
The small amplitude nearly-Gaussian fluctuations seed formation of large collapse events. The universal spatio-temporal form of these events together with the PDF for their maximum amplitudes
define the power-like tail of PDF for large amplitude fluctuations, i.e., the intermittency of strong turbulence.
\end{abstract}


\maketitle

\section{Introduction}

A  nonlinear Schr\"odinger equation (NLS)
\begin{equation}\label{nlstotal1}
 i\psi_t+\nabla^2\psi+\alpha|\psi|^2\psi+\beta|\psi|^{4}\psi=0,
\end{equation}
describes a wide class of interacting nonlinear waves and Bose-Einstein condensates. Here $t$ is the time, the Laplacian $\nabla^2$ is considered in the general dimension $D$, the constants $\alpha$ and
$\beta$ correspond to the cubic and quintic nonlinearities, respectively. Generally,   $\alpha |\psi|^2\psi$ gives the leading order nonlinear interaction. In many nonlinear systems $\alpha$ can vanish, which results in the
quintic nonlinear Schr\"odinger equation (QNLS)
\begin{equation}\label{nlsquintic1}
 i\psi_t+\nabla^2\psi+\beta|\psi|^{4}\psi=0.
\end{equation}
QNLS occurs e.g., in the Bose-Einstein condensate where the $s$-wave scattering length is set to zero by tuning Feshbach resonance \cite{PitaevskiiStringariBook2003,BrazhnyiKonotopPitaevskiiPRA2006}.
QNLS also occurs for general NLS type system near the transition from supercritical to subcritical bifurcations \cite{KuznetsovJETP1999,AgafontsevDiasKuznetsovJETPLett2008},  pattern formation
(in the context of quintic Ginzburg-Landau equation if $\beta$ is the complex constant) \cite{CrossHohenbergRevModPhys1993} and dissipative solitons (e.g., in lasers) \cite{SotoCrespoAkhmedievAnkiewiczPRL2000}.
Another possible experimental realization of equation (\ref{nlsquintic1}) is the optical pulse propagation in optical fiber using a nonlinear compensator of nonlinearity \cite{GabitovLushnikovOL2002}.

The standard cubic NLS (equation (\ref{nlstotal1}) with $\beta=0$) is integrable in dimension one (1D) by the inverse scattering transform \cite{ZakharovShabatJETP1972} with global existence of all solutions.
 In contrast,
  QNLS (\ref{nlsquintic1}) with positive real $\beta$ and any $D\ge 1$ can develop a finite time singularity (blow up)  such that  the amplitude of solution reaches infinity in a
finite time. The blow up is accompanied by
dramatic contraction of the function $\psi$ spatial extent,
that is called wave collapse or simply collapse \cite{VlasovPetrishchevTalanovRdiofiz1971,ZakharovJETP1972}. A sufficient condition for the collapse is
$H<0$, where
\begin{equation}\label{nls1}
 H=\int \big ( |\nabla
 \psi|^2-\frac{\beta}{3}|\psi|^{6}\big )d {\bf r}
\end{equation}
is the Hamiltonian (energy)  and the equation (\ref{nlsquintic1}) can be rewritten in the Hamiltonian form
\begin{eqnarray}
i\psi_t=\frac{\delta H}{\delta \psi^*}. \label{eq_Ham}
\end{eqnarray}

The case $D=1$, which we consider below, is critical  because any decrease of the power of nonlinearity in (\ref{nlsquintic1}) (i.e., replacement of $|\psi|^{4}\psi$ by $|\psi|^{4+\gamma}\psi, \ \gamma<0$) results in the global existence
of the solutions \cite{GinibreVeloJFunctAnal1979a,WeinsteinCommMathPhys1983,SulemSulem1999} for any real $\beta$.

Collapse of QNLS is not physical and  near singularity different physical regularization mechanisms come into play. These can be numerous nonlinear dissipation mechanisms such as inelastic collisions in
Bose-Einstein condensate which results in loss of particles from the condensate \cite{PitaevskiiStringariBook2003},
optical breakdown and formation of plasma in nonlinear optics \cite{BoydNonlinearOpticsBook2008}, or numerous non-dissipative regularization effects such as nonlinear saturation in laser-plasma
interactions \cite{LushnikovRosePRL2004}, different dispersive effects or non-paraxiality of optical beam (see e.g., \cite{FibichPapanicolaouSIAMJApplMath1999}).

In this paper we consider 1D QNLS with  linear and nonlinear dissipation so that QNLS (\ref{nlsquintic1})  is replaced by the following regularized QNLS (RQNLS):
\begin{eqnarray}\label{nlsregul}
 i\psi_t+(1-ia\epsilon)\partial_x^2\psi+(1+ic\epsilon)|\psi|^{4}\psi=i\epsilon \phi,
\end{eqnarray}
which can be also called as a complex quintic Ginzburg-Landau equation.
Here $x$ is the spatial coordinate replacing general ${\bf r}$, $0<\epsilon\ll 1$ is a small parameter so that to the leading
approximation $\epsilon\to 0$, QNLS (\ref{nlsquintic1}) is valid.
The coefficient $a\sim 1$ determines linear  dissipation and
the coefficient $c\sim 1$ is responsible for nonlinear
dissipation.
The linear dissipation has a viscosity-like form and can be resulted, e.g., from angular-dependent losses or the optical filtering \cite{RenningerChongWisePhysRevA2008,WainblatMalomedPhysD2009}.
The nonlinear dissipation in RQNLS corresponds to the three-photon absorption in optics \cite{BoydNonlinearOpticsBook2008} or four-body collisions which cause
loss of atoms from the Bose-Einstein condensate \cite{FerlainoKnoopEtAlPhysRevLett2009}.
The term $i\epsilon \phi$ describes the general forcing in the system.
The specific examples of the realization of RQNLS  (\ref{nlsregul}) are e.g., a propagation of light in a ring cavity with Kerr nonlinearity (see e.g., \cite{LushnikovSaffmanPRE2000})
or any quite general propagation of waves in nonlinear media with complex dispersion and nonlinear dissipation \cite{ZakharovLvovFalkovich1992}.

The right hand side (rhs) of equation (\ref{nlsregul}) provides forcing
and depends on the specific physical model. We consider two types of forcing. First is a deterministic forcing
\begin{eqnarray}\label{linearforcinggeneral}
\phi=\hat b\psi,  
\end{eqnarray}
which corresponds to a linear instability (amplification) in a system. Here $\hat b$ is the linear integral operator over $x$ such that its spatial Fourier transform $b_k$ is
the multiplication operator $\phi_k=b_k\psi_k$, where $b_k$ determines $k$-dependence of the amplification.
Below, if not mentioned otherwise, we implicitly assume the simplest case of $k$-independent amplification as $\hat b\psi=b\psi$ and thus, the equation (\ref{linearforcinggeneral}) takes the following form
\begin{eqnarray}\label{linearforcing}
\phi=b\psi,  
\end{eqnarray}
where $b$ is the positive constant.

The second type is a random additive forcing
\begin{eqnarray}\label{stochasticforcing}
\phi=\xi(t,x),
\end{eqnarray}
where zero in average stochastic term $\xi$  is a random Gaussian variable which is $\delta$-correlated in time  and has a finite spatial correlation such that
\begin{eqnarray}\label{chidef}
\langle \xi(t_1, x_1) \xi^*(t_2,
x_2)\rangle=\delta(t_1-t_2)\chi(|x_1-x_2|). \label{phi}
\end{eqnarray}
Here
\begin{eqnarray}\label{lcdef}
\chi(|x_1-x_2|)=b_g \exp{\left [-|x_1-x_2|/l_c \right ]},
\end{eqnarray}
$l_c$ is the correlation length of pump, $\langle \ldots \rangle$ denotes averaging over the statistics of $\xi$, and $b_g$ is the normalization constant.

\begin{figure}
\begin{center}
(a)
\includegraphics[width = 3.4 in]{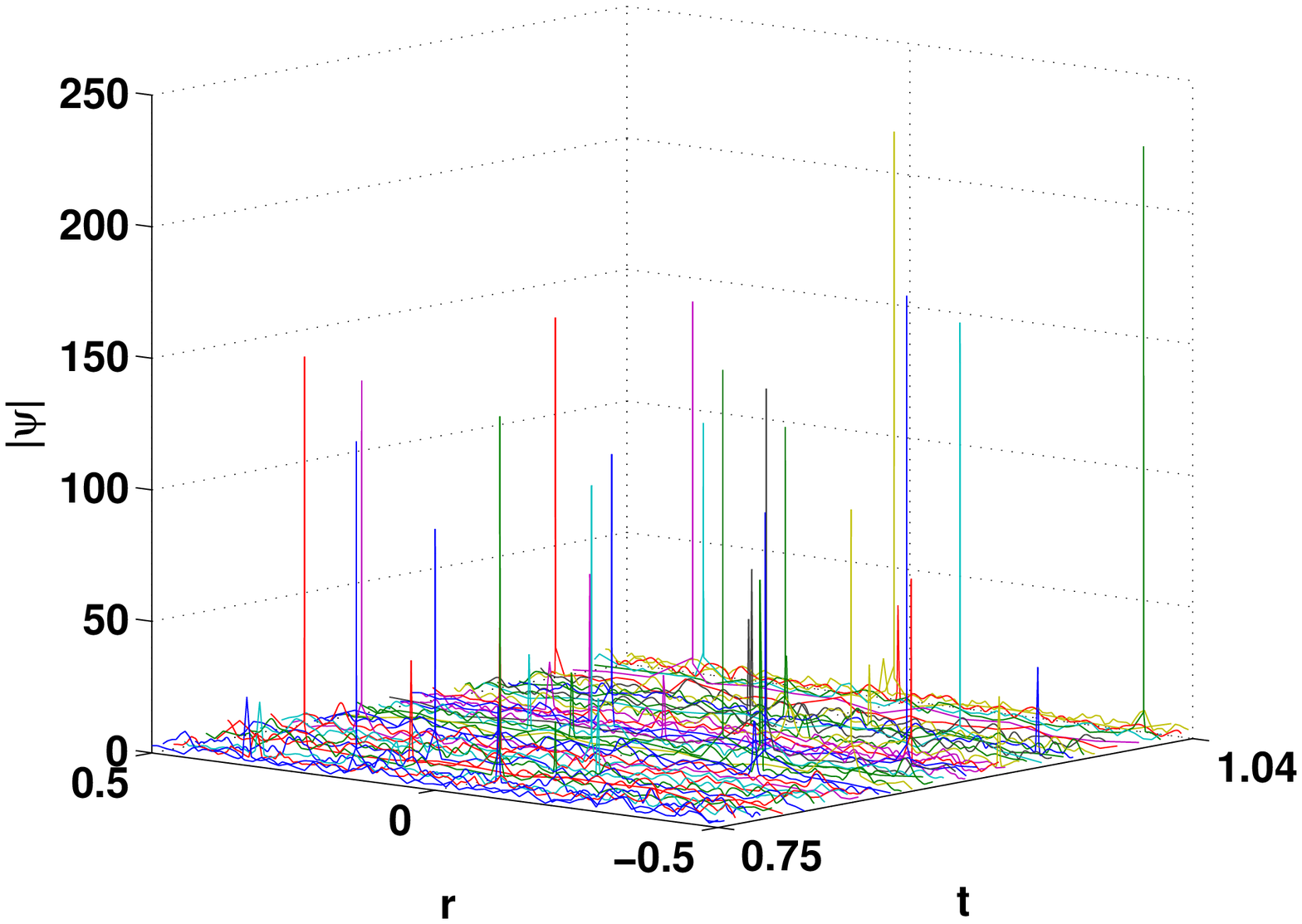}
(b)
\includegraphics[width = 3.4 in]{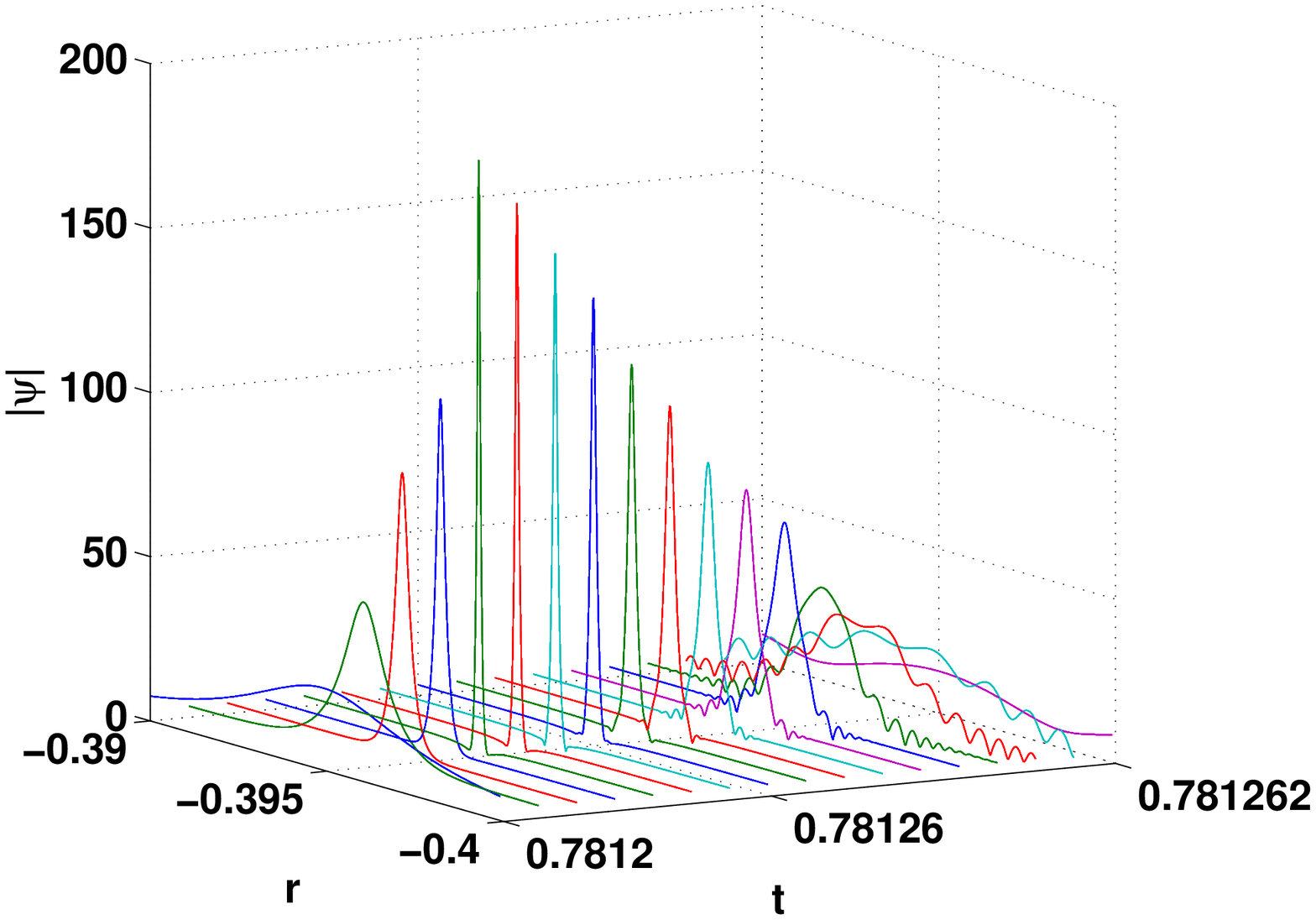}
\caption{(Color online) Spatial-temporal form of solution of RQNLS (\ref{nlsregul}) with the deterministic
forcing (\ref{linearforcing}), $b=10^4$, $\epsilon=2\times 10^{-3}$, and $a=c=1$ for different moments of time. (a) Collapses
(seen in plot as sharp peaks) occur at random spatial positions and
random times. (b) Zoom-in at evolution of single collapse event is shown
for the smaller time interval and the smaller spatial extent compare
with (a). It is seen that the collapse amplitude grows initially,
then goes through the maximum and finally decays due to dissipative
effects. Colors are added to (a) and (b) to help distinguish
different moments of time.} \label{fig:fig1}
\end{center}
\end{figure}

Forcing results in the pumping of energy into the system described by  RQNLS and subsequent formation of multiple collapse events randomly distributed in space and time as shown in Figure \ref{fig:fig1}a.
Figure \ref{fig:fig1}b shows a zoom-in into a temporal evolution of a spatial profile of a typical collapse event which involves growth  and subsequent decay of collapse amplitude.

\begin{figure}
\begin{center}
\vspace{2cm}
(a)
\includegraphics[width = 3.4 in]{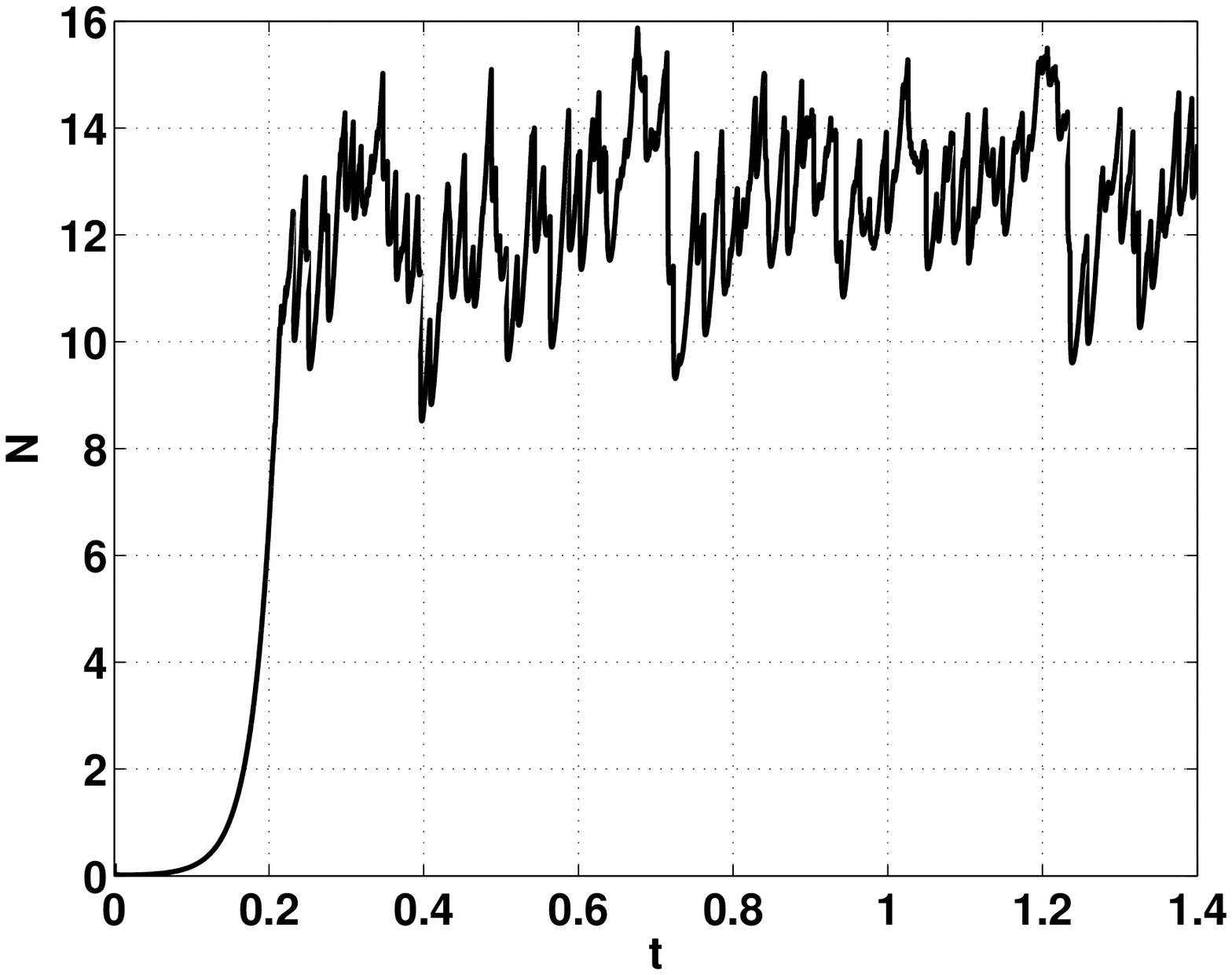}
(b)
\includegraphics[width = 3.4 in]{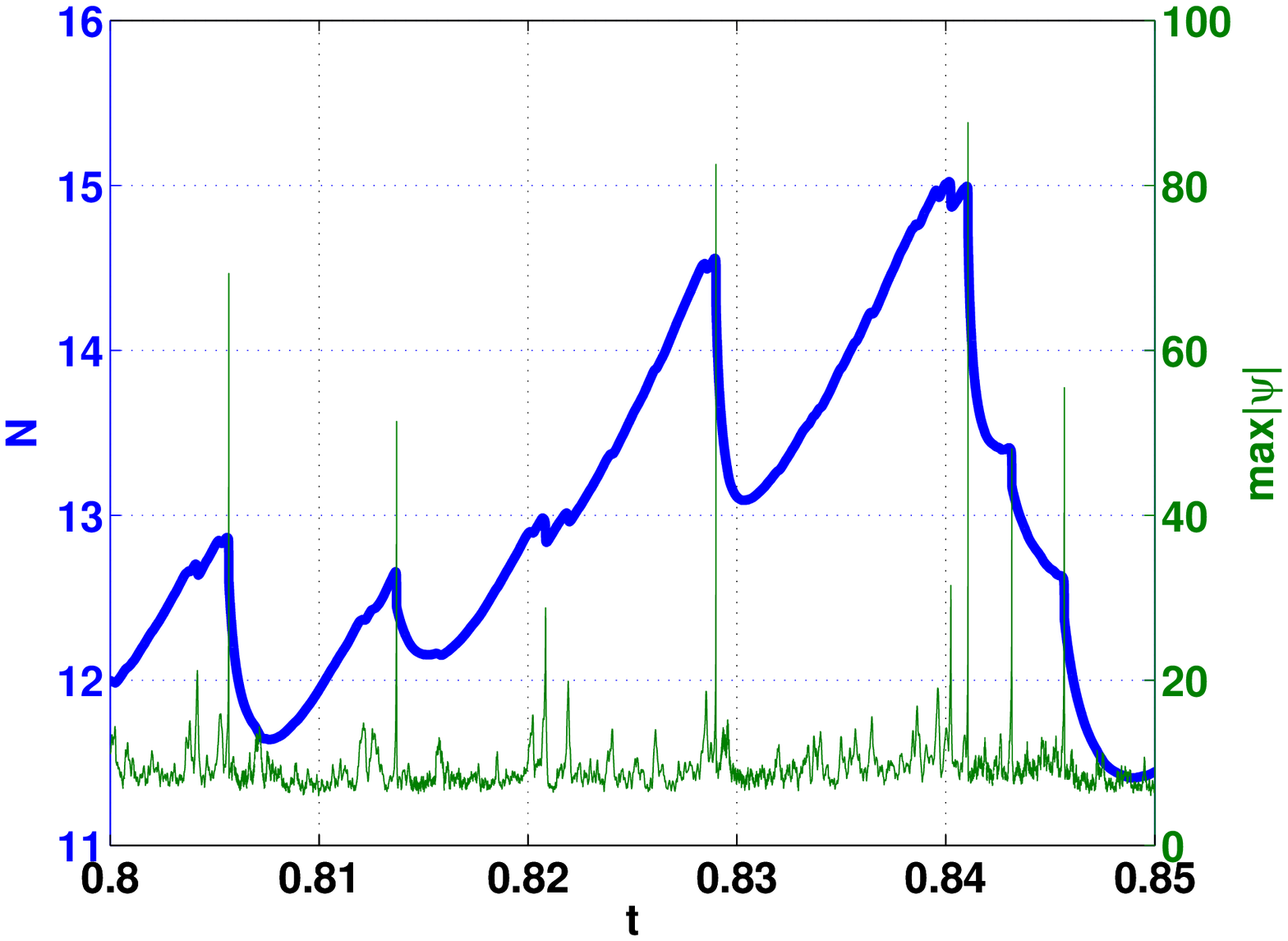}
\caption{(Color online) Time dependence of the number of particles $N$ for
numerical simulation of RQNLS (\ref{nlsregul}) with the deterministic forcing (\ref{linearforcing}) (left axis, thick solid line) superimposed
with the time evolution of maximum value of $|\psi|$ (right axis, solid line).
Parameters are $b=10^4, a=c=1, \epsilon=2\times 10^{-3}$ and the random small amplitude
initial condition is used. (a) Initially
$N$ grows until the statistical steady-state of developed turbulence is achieved so that $N$ remains constant in average over time. (b) Zoom-in at the smaller
scale in $t$. Solid thick curve (blue)
corresponds to $N(t)$. Solid thin curve (green) shows time dependence
of maximal spatial value of $|\psi|$. Scale for $N$ is on the left
vertical axis and scale for $\max |\psi|$ is on the right vertical
axis, respectively. It is seen that sharp decreases of $N$ is due to
the dissipation from large collapses. These decreases are compensated in
average by growth of $N$ from the linear forcing.} \label{fig:fig1N}
\end{center}
\end{figure}

After the initial transient, the solution of RQNLS achieves a statistical steady state  (i.e., state of the developed turbulence) as shown in Figure \ref{fig:fig1N}
through the time dependence of the following integral
\begin{equation}\label{Ndef}
N=\int |\psi|^2 d x,
\end{equation}
which has a meaning either of the number of particles in Bose-Einstein condensate, or an optical power (or sometimes energy) in optics as well as it is called by a wave action in oceanology
and many other nonlinear wave applications
\cite{ZakharovLvovFalkovich1992}. Below we refer to $N$ as the number of particles. It is seen in Figure \ref{fig:fig1N}b that the dissipation is important near large collapses while forcing works all time (because forcing is
$\propto N$). In the statistical steady state  the pumping of particles (forcing) in average is compensated by the dissipation which insures the state of the developed turbulence.

The focus of this paper is to describe the strong turbulence in RQNLS characterized by these nearly-singular collapse events.
By strong turbulence (we also call it  strong collapse turbulence) we mean the turbulence with strong non-Gaussian fluctuations as opposed to the weak turbulence with nearly
Gaussian fluctuations \cite{ZakharovLvovFalkovich1992}. The strong non-Gaussian fluctuations  are usually refer to as intermittency of turbulence \cite{FrischBook1995}.
The classical example of the strong turbulence is the Navier-Stokes turbulence. The old idea of the description of
  strong turbulence in the
Navier-Stokes equations through singularities of the
Euler equations still remains unsolved  \cite{FrischBook1995}. The forced Burgers equation
is a very rare example of an analytical description of
strong turbulence in which the tail of the probability density function (PDF) for negative
gradients follows a well-established $(-7/2)$ power law
\cite{EKhaninMazelSinaiPRL1997}, dominated by the spatio-temporal dynamics near formation of singular shocks (i.e., by pre-shocks).
The spectrum of the strong optical turbulence was considered for the two-dimensional cubic nonlinear Schr\"odinger equation in Ref.
\cite{DyachenkoNewellPushkarevZakharovPhysicaD1992} and power law tail  of PDF of the amplitude fluctuations was considered in Refs. \cite{ChungLushnikovVladimirovaAIP2009} and  \cite{LushnikovVladimirovaOptLett2010}.
Strong turbulence in RQNLS was first studied in Refs. \cite{IwasakiTohProgrTheorPhys1992} and \cite{TohIwasakiJPhysSocJap1992} with the $(-8)$  PDF power law scaling suggested in
Ref. \cite{IwasakiTohProgrTheorPhys1992}. Here we show that $(-8)$ is only an approximate scaling and it is determined by fluctuations of the waves at background which seed collapses as well as by the self-similar form of
collapsing solutions. We also found that the power law of PDF tail is only weakly sensitive to the type of forcing (linear amplification (\ref{linearforcing}) vs. additive random forcing (\ref{stochasticforcing}))
 for $\epsilon \ll 1$ showing the universal turbulent picture.




The paper is organized as follows. In Section
\ref{section:modulational} we consider the small amplitude fluctuations of the background of RQNLS turbulence, and show the universality of the spatial and temporal correlation functions
 and relate the correlation scale of these fluctuations to the scale of the modulational instability.
In Section \ref{section:standartownes} we review the collapsing self-similar solution of RQNLS for $\epsilon=0$ and establish the universality of the self-similar solution for $\epsilon\neq 0$
as a basic building block for large amplitude fluctuations. In Section \ref{section:PDFsimulations} we provide the results of numerical calculations of PDF for the amplitude fluctuations
and for the collapse maximums.    In Section \ref{section:PDFtails} we derive the analytical expression for the tail of PDF for amplitudes and compare with numerics.
In Section V we discuss the details of the numerical methods used. In Section
VI 
 the main results of the paper and
future directions are discussed.

\section{Modulational instability and  fluctuations of the background}
\label{section:modulational}

The forcing term in right hand side (rhs) of RQNLS (\ref{nlsregul}) pumps the number of particles $N$ into the system until the statistical steady state (also can be called by a developed turbulence state) is reached
as shown in Figure \ref{fig:fig1N}a
for the particular example of the deterministic forcing (\ref{linearforcing}). The system in developed turbulence state does not have memory of initial conditions because of the modulational instability.
That instability was derived for the cubic NLS \cite{BespalovTalanovJETPLett1966rus} (see also e.g. \cite{ZakharovLvovFalkovich1992}) but it is straightforward to generalize it for RQNLS as follows.
RQNLS (\ref{nlsregul}) with $\epsilon=0$ has a spatially uniform solution $\psi=\psi_0 \exp(i |\psi_0|^{4}t)$.
Linearization on the background of that solution in the form $\psi= \exp(i |\psi_0|^{4}t)\big [\psi_0+\delta\psi \exp({\nu t+i
kx})\big ]$ with $|\delta\psi/\psi_0|\ll 1$ gives the following instability growth rate  $\nu$   for the wavenumber $k$:
\begin{equation}\label{nu2}
\nu^2=k^2(4|\psi_0|^{4}-k^2).
\end{equation}
The instability occurs for $|k|<2|\psi_0|^{2}$ with the unstable branch $Re(\nu(k))=\nu(k)>0$ reaching maximum of $\nu(k)$ for
\begin{equation}\label{k2}
k_{max}^2=   2|\psi_0|^{4}.
\end{equation}
For $\epsilon \ne 0$, the expressions (\ref{nu2}) and (\ref{k2}) are still approximately valid provided $\nu(k_{max})=2|\psi_0|^4\gg 1/\tau_f$,
where $\tau_f$ is the typical time of forcing. For the deterministic forcing (\ref{linearforcing}), $\tau_f=(\epsilon b)^{-1}$ while for
the stochastic forcing  (\ref{stochasticforcing}), $\tau_f$ can be estimated from the condition that dissipation in average is compensated by the forcing. Then $\tau_f\sim \min[(\epsilon c p_0^4)^{-1},(\epsilon  a k_0^2)^{-1}]$,
where $p_0$ is the typical amplitude of the fluctuation of $|\psi|$ and $k_0$ is the typical wavevector which is estimated from (\ref{k2}) as $k_0\sim k_{max}$. Recalling now our assumption that $a\sim 1$ and $c\sim 1$
we arrive to a simpler estimate $\tau_f \sim (\epsilon  p_0^4)^{-1}$, e.g., using the parameters of Figure \ref{fig:fig1N} we obtain that $1/\tau_f\simeq 20 \ll \nu(k_{max})\simeq 3\cdot 10^2$.

\begin{figure}
\includegraphics[width = 3.4 in]{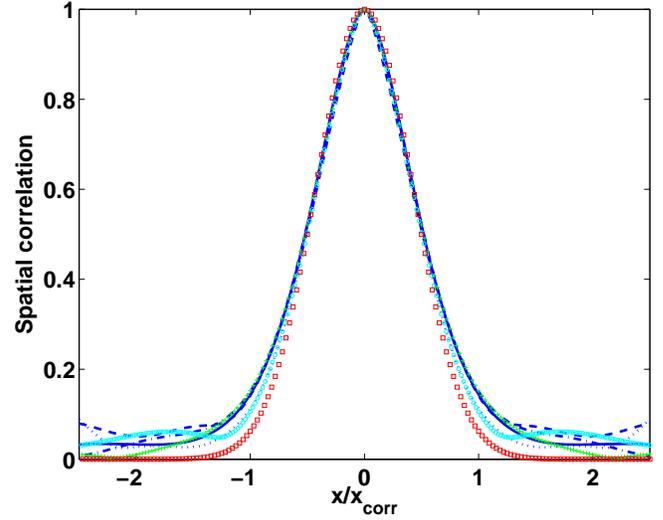} 
\caption{(Color online) Absolute value of the normalized spatial correlation function $|S(x)/S(0)|$ vs. $x/x_{corr}$. Parameters are $\epsilon= 5\cdot 10^{-3},a=c=1,b=10^4$ (solid blue line),
$\epsilon= 2\cdot 10^{-3},a=c=1,b=10^4$ (dashed-dotted blue line), $\epsilon= 10^{-3},a=c=1,b=10^4$ (dashed blue line), $\epsilon= 2\cdot 10^{-3},a=c=1,b=2\cdot 10^3$ (dotted blue line),
$\epsilon= 2\cdot 10^{-3},a=1,c=2,b=10^4$
 (green crosses), and $\epsilon = 2 \cdot 10^{-3},a=2,c=1,b=10^4$ (light blue circles). Red squares correspond to a Gaussian function $\exp{[-4\ln{2}(x/x_{corr})^2]}$ which has a unit FWHM.}
\label{corr_space_fwhm}
\end{figure}

Thus the dynamics of the background of turbulence can be characterized by the typical amplitude of the fluctuations $p_0$ and spatial scale $1/k_{max}=1/(2^{1/2}p_0^2)$. For simulations we define $p_0=(N/L_0)^{1/2},$
 where $L_0=\int dx$ is the computational
domain (without loss of generality we set $L_0=1$ in our simulations as described in Section V). We determine a correlation length $x_{corr}$ of $\psi$ through the full width at half maximum (FWHM) of  the spatial correlation function
$S(x)=\langle \psi(x_0,t)\psi^*(x_0+x)\rangle$. Here, $\langle \ldots \rangle$ denotes averaging over space and time which is assumed by the ergodicity to give the same result
as the average over the ensemble of simulations.
Figure \ref{corr_space_fwhm} shows $|S(x)/S(0)|$ vs. $x/x_{corr}$ for a set of simulations with  different sets of parameters. Each curve is calculated in the statistical steady state.
It is seen that $|S(x)/S(0)|$ is well approximated by the universal function of
$x/x_{corr}$ which is close to a Gaussian function $\exp{[-(x/x_0)^2]}$ while $S(0)=p_0^2$ and $x_0$ is chosen from the condition to have  the same FWHM of the Gaussian function and $|S(x)/S(0)|$.
It implies $x_0/x_{corr}= 1/(2\sqrt{\ln{2}}).$
Note with the increase of the ensemble of simulations (i.e., the simulation time used to calculate $S(x)$),
the value of $\mathrm{Im}(S(x))$ approaches to $0$. 

\begin{figure}
\includegraphics[width = 3.4 in]{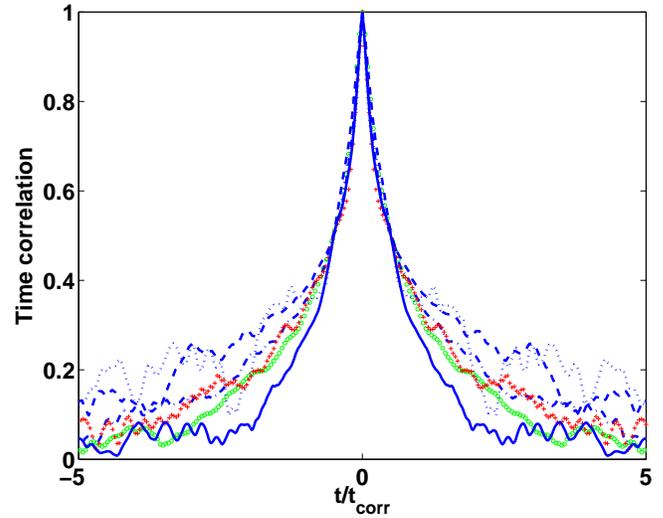} 
\caption{(Color online) Absolute value of the normalized temporal  correlation function $|P(t)/P(0)|$ vs. $t/t_{corr}$. Parameters are $\epsilon= 5\cdot 10^{-3},a=c=1,b=10^4$ (green circles),
$\epsilon= 2\cdot 10^{-3},a=c=1,b=10^4$ (dashed-dotted  blue line), $\epsilon= 10^{-3},a=c=1,b=10^4$ (dashed blue line), $\epsilon= 2\cdot 10^{-3},a=c=1,b=2\cdot 10^3$ (dotted blue line),
$\epsilon= 2\cdot 10^{-3},a=1,c=2,b=10^4$ (red crosses), and $\epsilon = 2 \cdot 10^{-3},a=2,c=1,b=10^4$ (solid blue line). }
\label{corr_time}
\end{figure}

We also determine a correlation time $t_{corr}$ of $\psi$ through FWHM of the temporal correlation function $P(t)=\langle \psi(x,t_0)\psi^*(x,t_0+t)\rangle$. Figure \ref{corr_time} shows  $|P(t)/P(0)|$  as a function of normalized time  $t/t_{corr}$ calculated in the statistical steady state.
It is seen that for different set of parameters, $|P(t)/P(0)|$ is well approximated by the universal function of
$t/t_{corr}$. The fluctuations in the tails of $|P(t)/P(0)|$ in Figure \ref{corr_time} are due to the finite size of the statistical ensemble.

\begin{figure}
\includegraphics[width = 3.4 in]{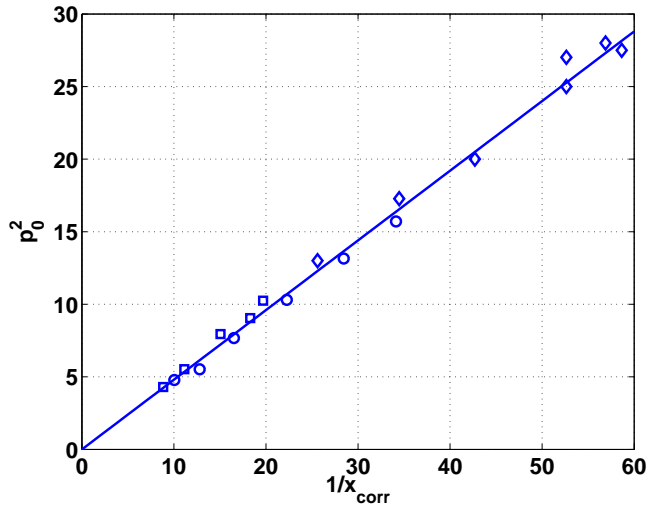} 
\caption{(Color online) Dependence of $p_0^2$ on $x_{corr}^{-1}$, 
where $x_{corr}$ is the FWHM of spatial correlation function $S(x)$.
Circles correspond to the case of $k$-independent deterministic forcing with $a=1,b=10^3,c=1,\epsilon=2\cdot 10^{-3}$; $a=1,b=2\cdot 10^3,c=1,\epsilon=2\cdot 10^{-3}$; $a=1,b=3 \cdot 10^3,c=1,\epsilon=2\cdot 10^{-3}$; $a=2,b=10^4,
c=1,\epsilon=2\cdot 10^{-3}$; $a=1,b=10^4,c=2,\epsilon=2\cdot 10^{-3}$; $a=1,b=10^4,c=1,\epsilon=5\cdot 10^{-3}$, from leftmost to rightmost.
Squares are for the ``$k$-limited" deterministic forcing  (\ref{linearforcinggeneral}),  (\ref{bkdefcutoff}) with $k_{cutoff} = 10\pi$ for $a=1,b=10^3,c=1,\epsilon=2\cdot 10^{-3}$; $a=1,b=3 \cdot 10^3,c=1,\epsilon=2\cdot 10^{-3}$; $a=2,b=10^4,
c=1,\epsilon=2\cdot 10^{-3}$; $a=1,b=10^4,c=2,\epsilon=2\cdot 10^{-3}$; $a=1,b=10^4,c=1,\epsilon=2\cdot 10^{-3}$, from leftmost to rightmost.
Diamonds are for the random forcing with $a=1,b_g= 250,c=1,\epsilon=5\cdot 10^{-3}$; $a=1,b_g = 1250,c=1,\epsilon=5\cdot 10^{-3}$; $a=1,b_g= 1250, c=1,\epsilon=2\cdot 10^{-3}$;
$a=1,b_g=2500,c=1,\epsilon=2\cdot 10^{-3}$; $a=2,b_g= 5000,c=1,\epsilon=2\cdot 10^{-3}$;$a=1,b_g=5000,c=2,\epsilon=2\cdot 10^{-3}$; $a=1,b_g=5000,c=1,\epsilon=2\cdot 10^{-3}$, and $l_c=0.02$ in all cases from leftmost to rightmost. Solid line has a slope $0.48$ in accordance with (\ref{p02xcorr}).  }
\label{N_corr}
\end{figure}

Figure \ref{N_corr} shows that the dependence of $p_0(x_{corr})$ is well approximated by
\begin{equation}\label{p02xcorr}
p_0^2 x_{corr}=Const\simeq 0.48
\end{equation}
for different values of parameters of RQNLS and  for 3 different types of forcing.
This gives another indication that the modulational instability determines the correlation length $x_{corr}$
through the amplitude of the background fluctuations $p_0$ in agreement with the  equation (\ref{k2}).

Here, the first type is the standard forcing (\ref{linearforcing})
which, while not introducing
any scale by itself (it pumps energy into all Fourier modes $k$), creates the scale $x_{corr}$ indirectly through the development of the modulational instability.
Circles in Figure \ref{N_corr}  correspond to that type of forcing.

\begin{figure}[ptb]
\includegraphics[width = 3.4 in]{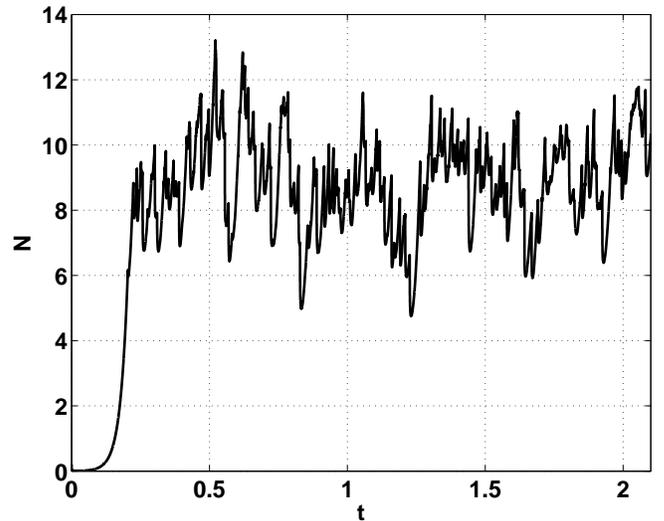} 
\caption{Time dependence of the number of particles $N$ in the case of ``$k$-limited" deterministic forcing  (\ref{linearforcinggeneral}),(\ref{bkdefcutoff}) with  $k_{cutoff}=10\pi$.
Parameters are $a=c=1, b_0=10^4, \epsilon=2\cdot 10^{-3}$.}
\label{force10m_l2}
\end{figure}

The second type  is a particular example of the general type of forcing
(\ref{linearforcinggeneral}). We define it by introducing a cutoff wave number $k_{cutoff}$ for the amplification:
\begin{equation}\label{bkdefcutoff}
b_k= b_0\neq 0,  \ \
|k|\le k_{cutoff};  \quad b_k=0, \ \  |k|>k_{cutoff}.
\end{equation}
In Figures  \ref{N_corr},\ref{force10m_l2}, for this second type of forcing we use  (\ref{linearforcinggeneral}) and  (\ref{bkdefcutoff}) with $k_{cutoff}=10\pi$ which means that only 11 modes are amplified.
The resulting $N(t)$ dependence in Figure \ref{force10m_l2} is similar to Figure
\ref{fig:fig1N}. In simulation of Figure \ref{force10m_l2}, $k_{max}\simeq 2p_0^2\simeq  20 \sim k_{max}$, and thus, $k_{cutoff}\sim k_{max}$.
Squares in Figure \ref{N_corr} correspond to this type of forcing.

\begin{figure}
\includegraphics[width = 3.4 in]{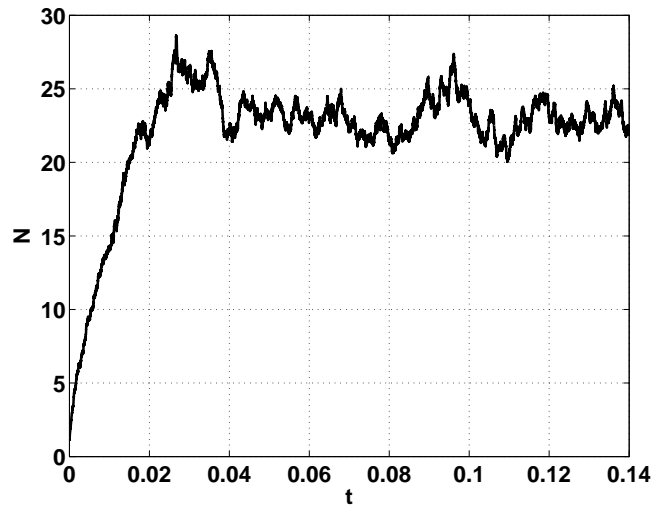} 
\caption{Time dependence of the number of particles $N$ in the case of random forcing. Parameters are $a=c=1,\epsilon=10^{-3},l_c=0.02,b_g=12500$.}
\label{l2_random}
\end{figure}

The third type is a random additive forcing (\ref{stochasticforcing}).
Figure \ref{l2_random} shows $N(t)$ dependence from the simulations with a random forcing and the result is similar to Figures \ref{fig:fig1N} and \ref{force10m_l2}. Diamonds in Figure \ref{N_corr} correspond to this type of forcing.

We conclude that the turbulence in RQNLS is independent of the type of forcing used (up to the normalization). The own scale of the forcing (if exist) is not important and the correlation length
is determined by the modulational instability scale.
Qualitative picture is the following: the interplay between forcing and dissipation in RQNLS defines
the amplitude $p_0$ and the correlation length $x_{corr}$ to be related  by  (\ref{p02xcorr}).

We also note that the principle difference between RQNLS turbulence  and
Navier-Stokes turbulence is the absence for RQNLS of well-defined inertial interval at which both forcing and dissipation are not important.
For RQNLS dissipation and forcing generally act at all scales.
Thus study of the spectrum of $\langle |\psi_k|^2\rangle$ could give
much less information compare with e.g., weak turbulence
\cite{ZakharovLvovFalkovich1992}.
Instead below we focus on the study of the tail of PDF for large
fluctuations.

\section{Collapse and its regularization in RQNLS }
\label{section:standartownes}

Fluctuations of background result in multiple formations of collapses in RQNLS as shown in Figure \ref{fig:fig1}.
For $\epsilon=0$, the collapsing solution of RQNLS has the following self-similar form
\begin{eqnarray}\label{selfsimilar}
&\psi(r,t)=\frac{1}{L^{1/2}}V(\rho,\tau)e^{i\tau +iL L_t
\rho^2/4},\\
&\rho=\frac{r}{L}, \quad \tau=\int^t \frac{dt'}{L^2(t')}, \nonumber
\end{eqnarray}
where $L=L(t)$ is the dynamically evolving spatial scale of the
collapsing solution at a given moment of time, $\rho$ and $\tau$ are sometimes called by blow up variables.
$V(\rho,\tau)$ is well approximated for small $L\ll 1$ by the ground state soliton solution $R_0(\rho):$ $V(\rho,\tau)\simeq R_0(\rho)$. Here
$R_0(x)$ is the positive-definite solution of the equation $-R_0+\partial^2_xR_0+R^3_0=0$, which follows from RQNLS assuming that $\psi=e^{i t}R_0(x)$ and $\epsilon=0.$
 The explicit expression for $R_0(x)$ is given by
\begin{eqnarray} \label{R1D}
R_0(x)=\frac{3^{1/4}}{(\cosh{ 2x})^{1/2}}.
\end{eqnarray}
The Hamiltonian (\ref{eq_Ham}) ($\beta=1$ according to (\ref{nlsregul})) vanishes at the  ground state soliton solution (\ref{R1D}). The number of particles (\ref{Ndef}) at (\ref{R1D}) is given by
\begin{eqnarray} \label{Nc}
N_c=\frac{\sqrt{3}\pi}{2}.
\end{eqnarray}
$N_c$ determines the boundary between collapsing and non-collapsing solutions: collapse is impossible for $N<N_c.$
Note that the ground state soliton solution for NLS (\ref{nlstotal1}) with $\beta=0$ in dimension two (2D) is often referred to as the Townes solution and it plays a similar role to (\ref{R1D}) in the collapse
of 2D NLS.
As $L(t)$ decreases with $t\to t_0$, the spatial distribution of $V(\rho)\to R(\rho)$ for $\rho\lesssim 1$, where $t_0$ is the collapse time (time at which a singularity develops in the solution of RQNLS with $\epsilon=0$).
It implies that the number of particles in the collapsing region, $N_{collpase}\to N_c$ as  $t\to t_0$, i.e., the collapse of QNLS is strong one as opposed to the weak collapse
\cite{ZakharovKuznetsovJETP1986,SulemSulem1999} for which the number of particles in the
collapsing region vanishes as $t\to t_0$.

The leading order behavior of $L(t)$ for $t\to t_0$ can be estimated from the scaling analysis as  $L(t)\sim (t_0-t)^{1/2}$ but the criticality of 1D QNLS results in the following log-log modification of that scaling
\cite{FraimanJETP1985,LandmanPapanicolaoSulemSulemPRA1988,LeMesurierPapanicolaouSulemSulemPhysicaD1988b,DyachenkoNewellPushkarevZakharovPhysicaD1992}
\begin{eqnarray}
L\simeq \left(\frac{2\pi(t_0-t)}{\ln\ln[1/(t_0-t)]}\right)^{1/2}.
\label{double}
\end{eqnarray}

\begin{figure}
\begin{center}
(a) \includegraphics[width = 3.4 in]{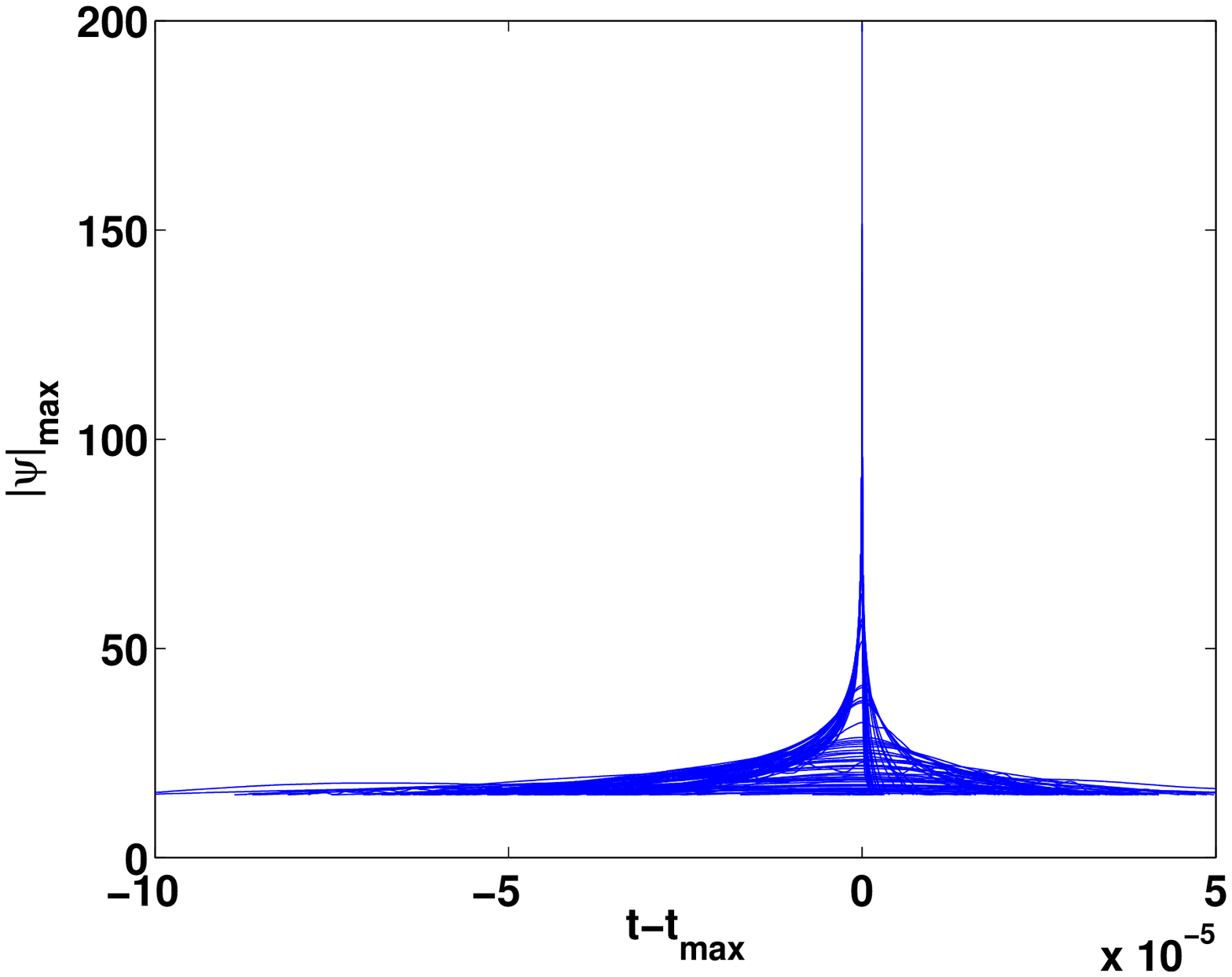} 
(b) \includegraphics[width = 3.4 in]{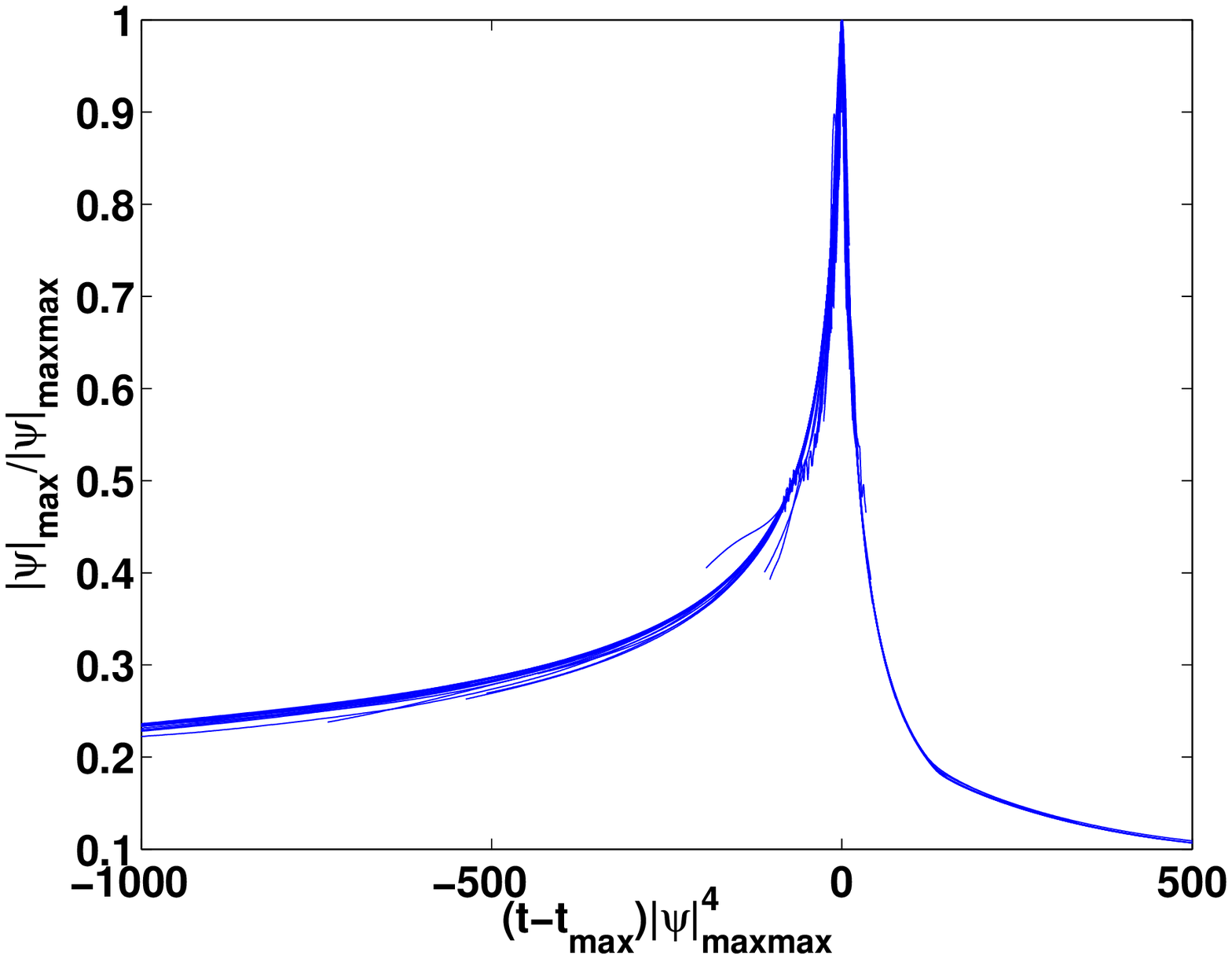} 
\caption{(Color online) (a) Dependence of $|\psi|_{max}$ on $(t-t_{max})$. 75 individual collapse events are collected for $|\psi|_{max} \geq 15$ with parameters $a=1,b=10^4,c=1, \epsilon= 2\cdot 10^{-3}$. (b) The same dependence as in (a) but in rescaled units
$ (t-t_{max})|\psi|_{maxmax}^4$  and $|\psi|_{max}/|\psi|_{maxmax}$. The time intervals over which $|\psi|_{max}/|\psi|_{maxmax}$ are plotted correspond to the time intervals of (a) (i.e., temporal interval is fixed in non-rescaled units).
 As a result, collapses with relatively small $|\psi|_{maxmax}$ extends over small intervals of $ (t-t_{max})|\psi|_{maxmax}^4$ in (b).}
\label{coll_over}
\end{center}
\end{figure}

\begin{figure}
\begin{center}
(a)\includegraphics[width = 3.0 in]{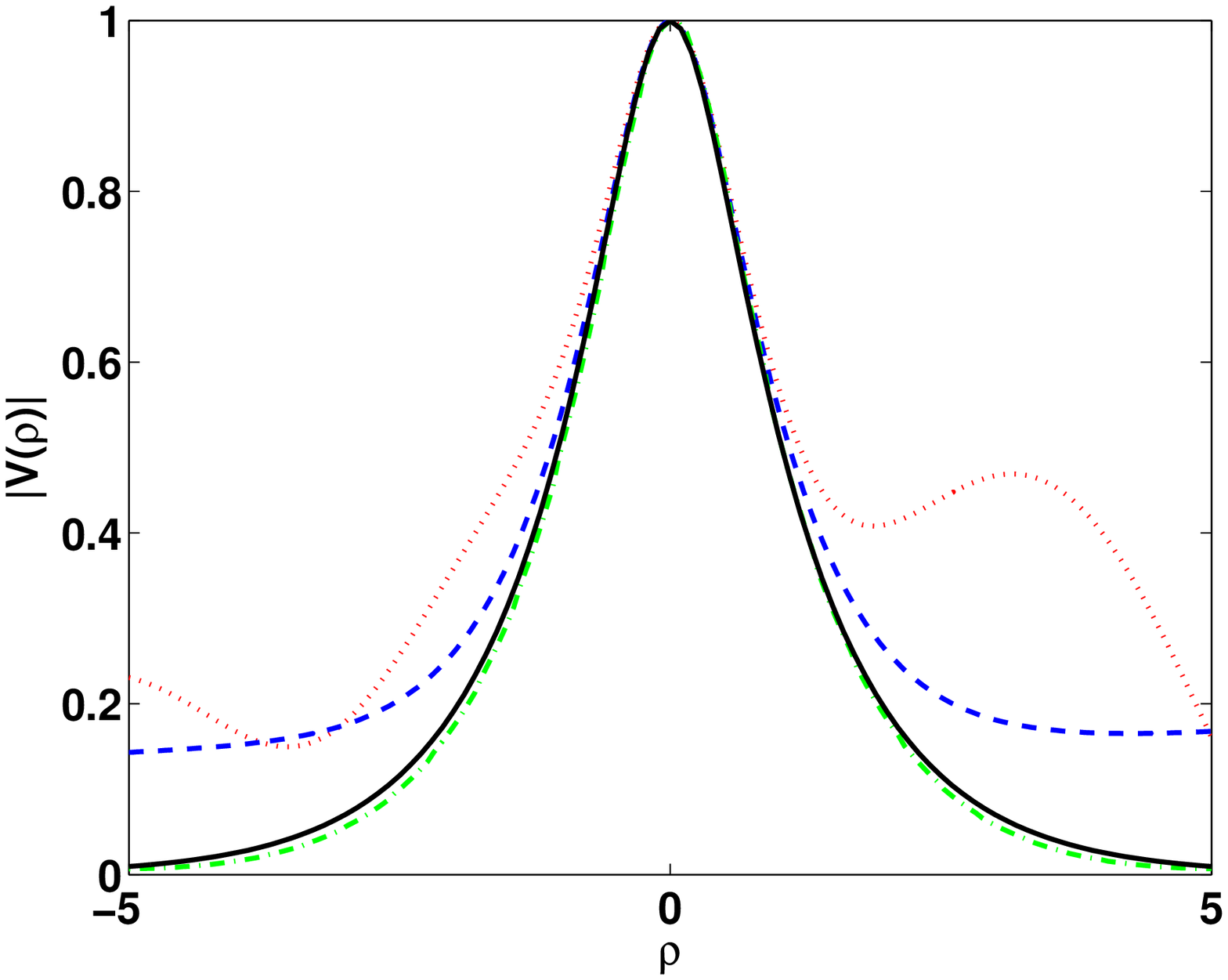} 
(b)\includegraphics[width = 3.0 in]{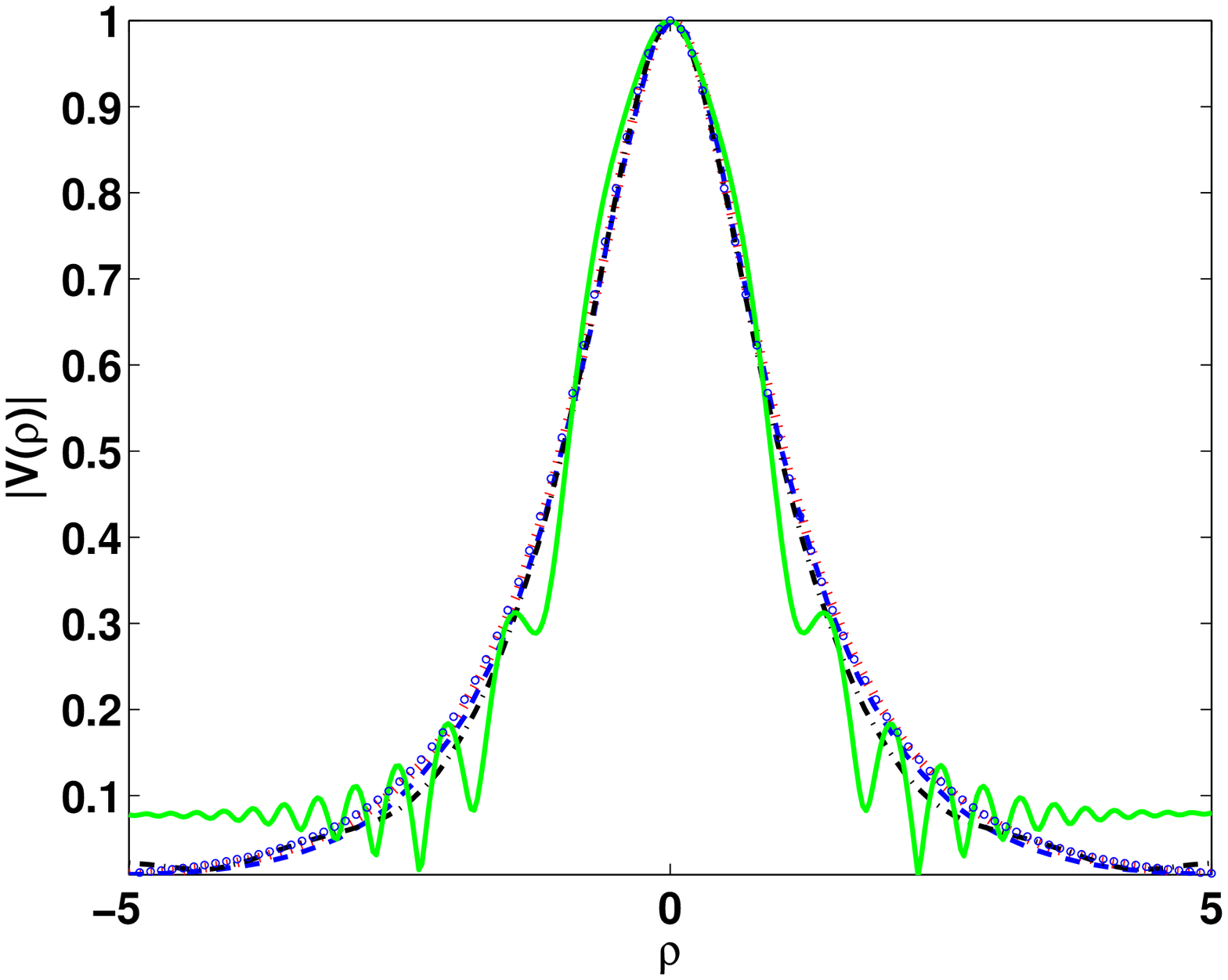} 
(c)\includegraphics[width = 3.0 in]{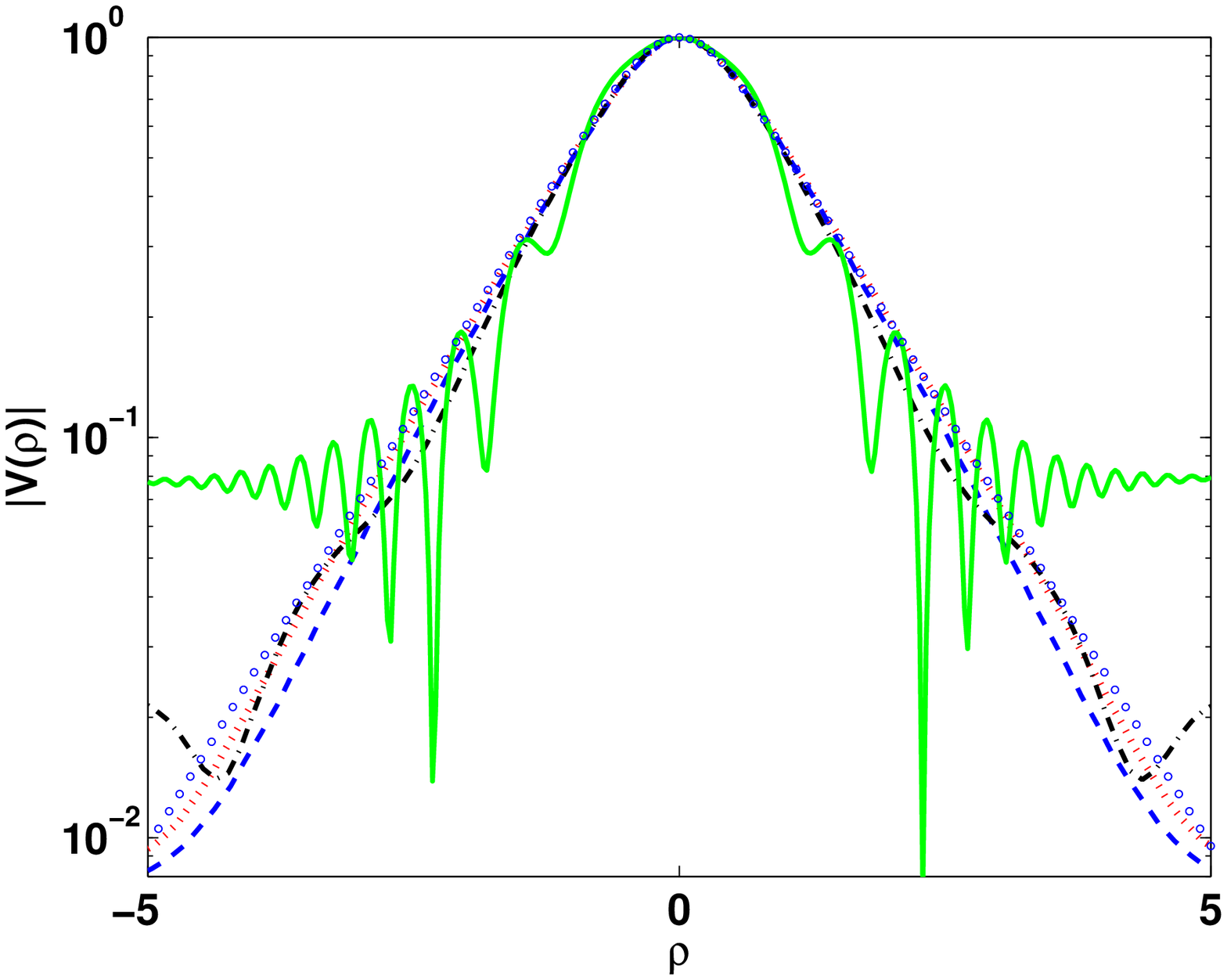} 
\caption{(Color online) Scaled profiles of numerical solution $|\psi(x)|$ in the vicinity of $t=t_{max}$ at which $|\psi|_{max}$ reaches the maximum $|\psi|_{maxmax}$.
Parameters are  $\epsilon =2\cdot 10^{-3},a=c=1,b=10^4$. (a) Scaled numerical solution $|\psi(x)|$ at $t=t_{max}-(4\cdot 10^{-5})$ (dotted red), $t = t_{max}- 10^{-5}$ (dashed blue), and $t = t_{max}$ (dashed-dotted green).
Solid black line shows  the normalized  ground state soliton  $\tilde{R}(\rho)=3^{-1/4}R_0(\rho)$ (\ref{R1D}).
(b) Scaled numerical solution $|\psi(x)|$ at $t=t_{max}-10^{-8}$ (dotted red), $t=t_{max}$ (dashed blue), $t=t_{max}+10^{-8}$ (dashed-dotted black), $t=t_{max}+(7\cdot 10^{-8})$ (solid green). Circles show the normalized  ground state soliton as in (a).
(c) Logarithmic scale of Figure (b).
}
\label{rho}
\end{center}
\end{figure}

RQNLS with $\epsilon>0$ does not allow singular collapses. Instead, for $0<\epsilon\ll 1$, the collapse amplitude $|\psi|_{max}(t)\equiv \max\limits_{x}|\psi(x,t)|$ goes (as a function of time) through a maximum
$|\psi|_{maxmax}=\max\limits_{t}|\psi|_{max}(t)$ at some $t=t_{max}$ and decays after that as shown in Figure \ref{coll_over}a.
The spatial form of the collapsing solution is still well approximated  by (\ref{selfsimilar}) and (\ref{R1D})  before and shortly after $t=t_{max}$ as seen in Figures \ref{rho}a,b and c.
It is seen in Figure \ref{rho}a that with the growth of $|\psi|_{max}(t)$ the normalized shape of solution approaches to (\ref{R1D}).
But with the decay of $|\psi|_{max}(t)$ for $t>t_{max}$, the  normalized shape of solution departs from (\ref{R1D}) and produces growing oscillating tails  as seen in Figures \ref{rho}b and \ref{rho}c.
In the vicinity of  $t=t_{max}$ the forcing on the rhs of
RQNLS (\ref{nlsregul}) can be neglected. The resulting
equation can be written in rescaled units $ t|\psi|_{maxmax}^4,$ $
x|\psi|_{maxmax}^2$, and $ \psi/|\psi|_{maxmax}$ to have exactly the same form, i.e., RQNLS without forcing is invariant with respect to these scaling transformations.
As shown in Figure \ref{coll_over}b vs. Figure \ref{coll_over}a,
$|\psi|_{max}(t)$ rescaled in these units exhibits a universal behavior: all curves collapse on a single curve
in the neighborhood of large collapses. That universality is
independent of the complicated structure of optical turbulence. We conclude up to rescaling all collapse events in RQNLS are identical which is qualitatively similar to the universality of collapse in QNLS. This
universality is a characteristic feature of the dissipative terms
in  RQNLS  (\ref{nlsregul}).

\begin{figure}
\includegraphics[width = 3.4 in]{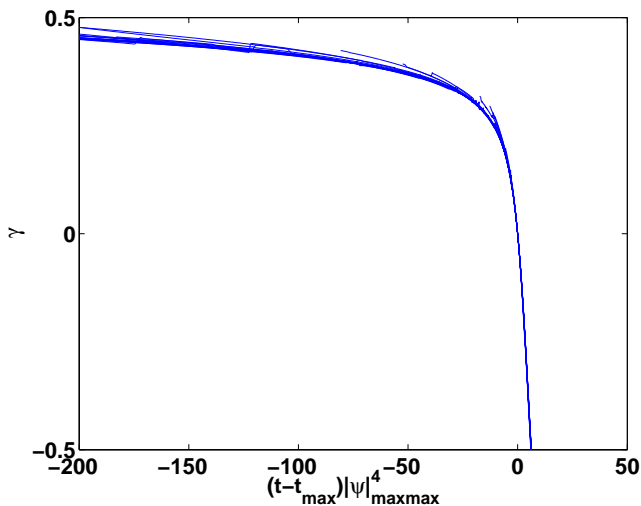} 
\caption{(Color online) Dependence of $\gamma(t) = -L_tL$ on $(t-t_{max})|\psi|_{maxmax}^4$. The same individual collapse events as in Figure \ref{coll_over} are collected for $|\psi|_{max} \geq 30$ in non-rescaled units.}
\label{gamma}
\end{figure}

A function
\begin{equation}\label{gammadef}
\gamma\equiv -L\frac{dL}{dt},
\end{equation}
changes slowly with $t$ compared to $L$ at $t\lesssim t_{max}$ except very small neighborhood of $t=t_{max}$ as shown in Figure \ref{gamma}. The dependence of $\gamma(t)$ is shown in Figure \ref{gamma} in the rescale time unit
$ (t-t_{max})|\psi|_{maxmax}^4$ similar to Figure \ref{coll_over}b. We determine  $L(t)$ numerically from individual collapse events with $|\psi|_{maxmax}>30$
using (\ref{selfsimilar}) and (\ref{R1D}) as $L(t)=3^{1/2}/|\psi|_{max}(t)^2$.
Thus, the universality of each collapse event is seen for $\gamma(t)$ also.

We conclude in this Section that the collapse events in RQNLS turbulence are all universal ones after a proper rescaling.

\section{PDF of $|\psi|$}
\label{section:PDF}

\subsection{PDFs from simulations}
\label{section:PDFsimulations}

\begin{figure}
\begin{center}
 (a)
\includegraphics[width = 3.4 in]{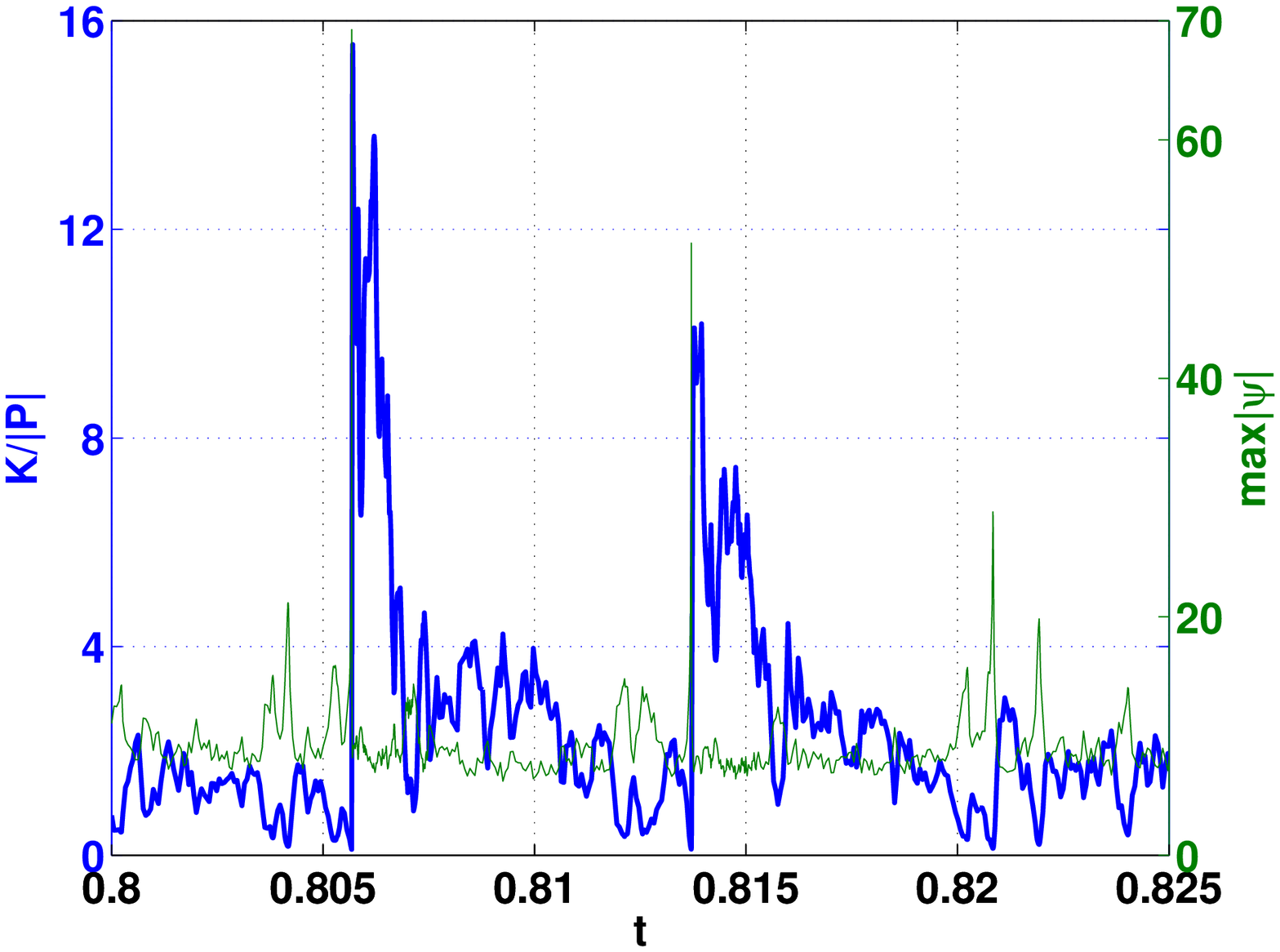}
(b)
\includegraphics[width = 3.4 in]{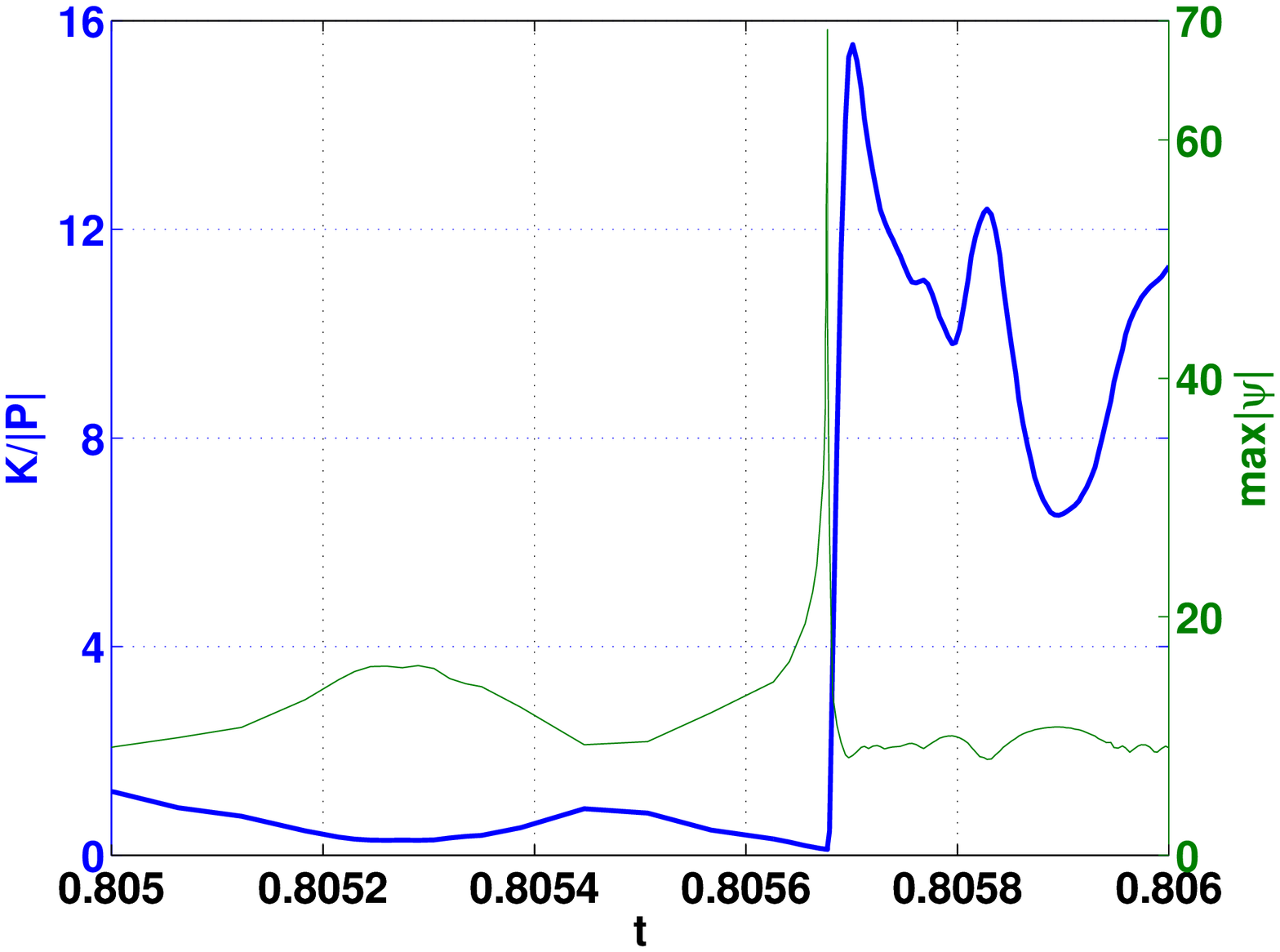}
\caption{(Color online) (a) Time dependence of $\frac{K}{|P|}$, where $K$ is the kinetic energy and $P$ is the potential energy defined in the text.
Parameters are $a=1,b=10^4,c=1, \epsilon=2\cdot 10^{-3}$.
Thick solid curve (blue) corresponds to $\frac{K}{|P|}$. Thin solid curve (green) is time dependence of maximal spatial value of $|\psi|$.
Scale for $\frac{K}{|P|}$ is on the left vertical axis and scale for $\max|\psi|$ is on the right vertical axis. (b) Zoom-in at the same curves as in (a) for a single collapse event.}
\label{kin_pot}
\end{center}
\end{figure}

Once the amplitude of a collapse event $|\psi|_{max}$ reaches the maximum $|\psi|_{maxmax}$, it starts  decreasing and subsequently, the collapse event decays into outgoing  waves as seen in Figure
 \ref{rho}b and  \ref{rho}c. Decaying waves are almost liner ones as can  be seen from the time dependence of the ratio of the kinetic energy $K=\int |\partial_x\psi|^2 d x$ and the potential energy
 $P=-\frac{1}{3}\int |\psi|^6 d x$  in Figure \ref{kin_pot}a. Note that  the Hamiltonian (\ref{nls1}) is a sum of $K$ and $P$ and during the growth of the amplitude $|\psi|_{max}$ of collapse
 event, these terms nearly cancel each other
 if the integrals are calculated only inside the collapsing region $\rho \lesssim 1.$
 While at the decay phase of the collapse event, the amplitude that balances is strongly violated in favor of $K$ meaning that solution is becoming nearly linear one.
Figure \ref{kin_pot}b shows zoom-in at a single collapse event. It is seen that while amplitude of collapse event decays, the kinetic energy from outgoing waves dominates over potential energy.
Superposition of many of these almost linear waves from multiple collapse event  forms a
nearly random Gaussian field by the central limit theorem \cite{Gardiner2004}. That random field seeds new collapse events.

\begin{figure}
\begin{center}
(a)
\includegraphics[width = 3.0 in]{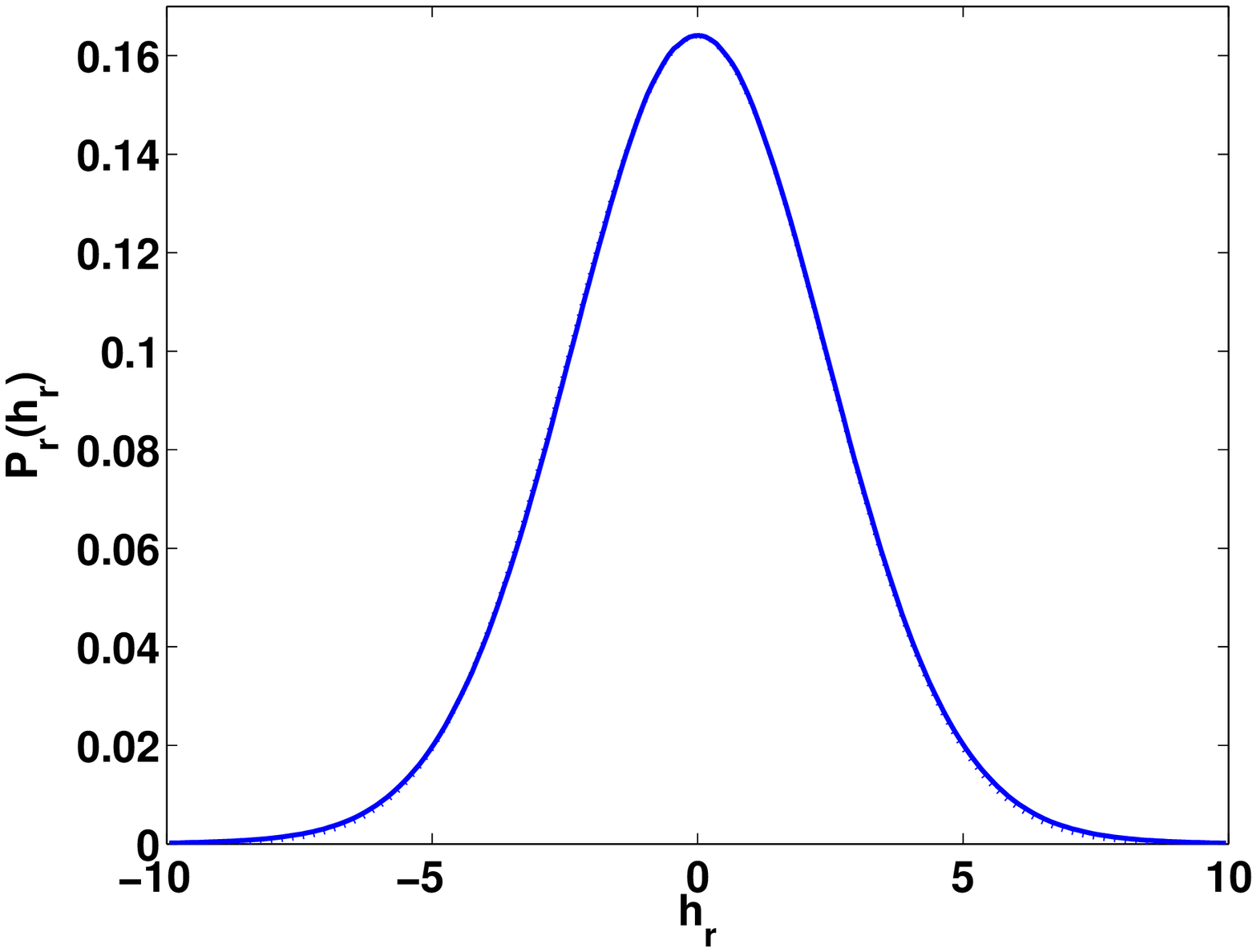} 
(b)
\includegraphics[width = 3.0 in]{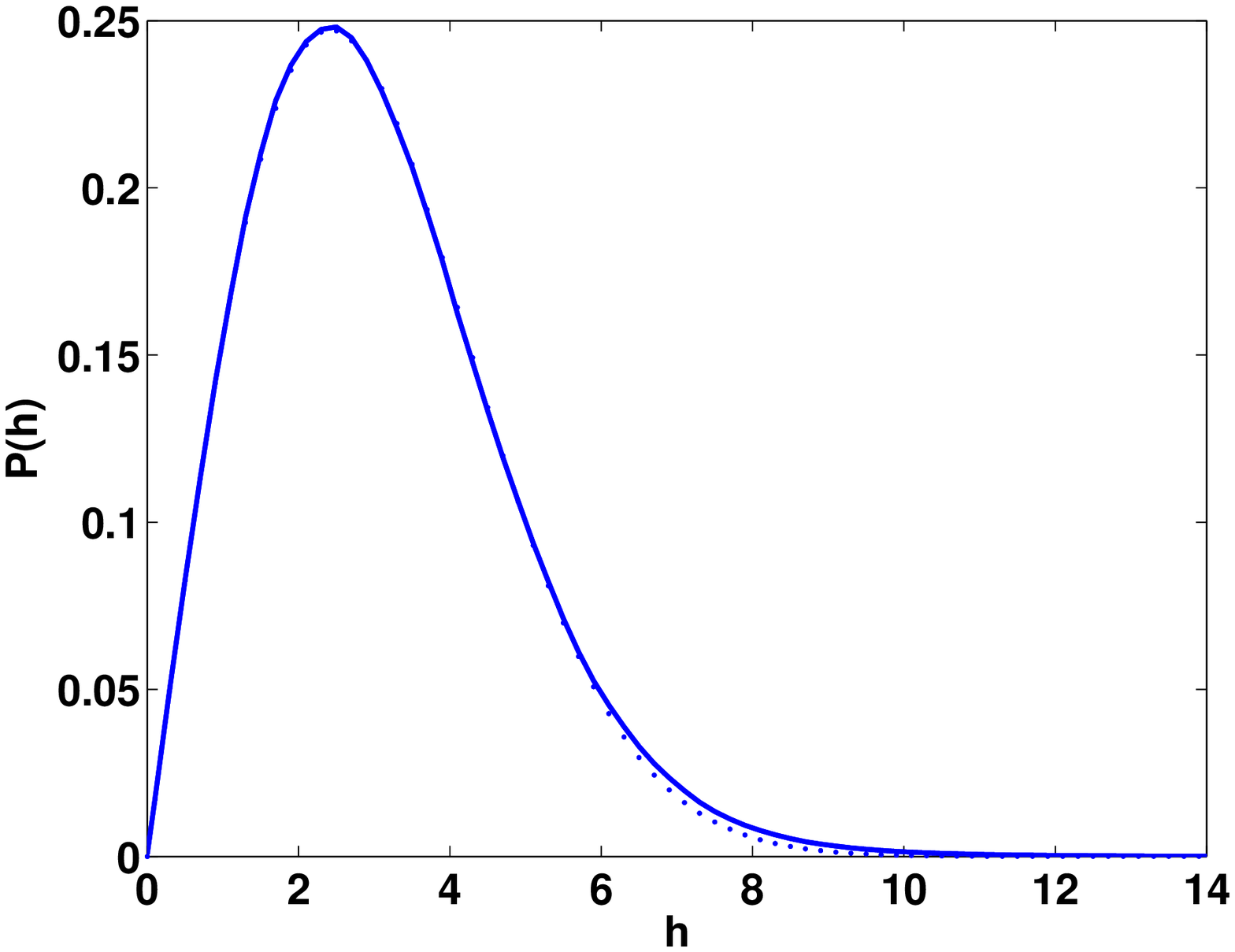} 
(c)
\includegraphics[width = 3.0 in]{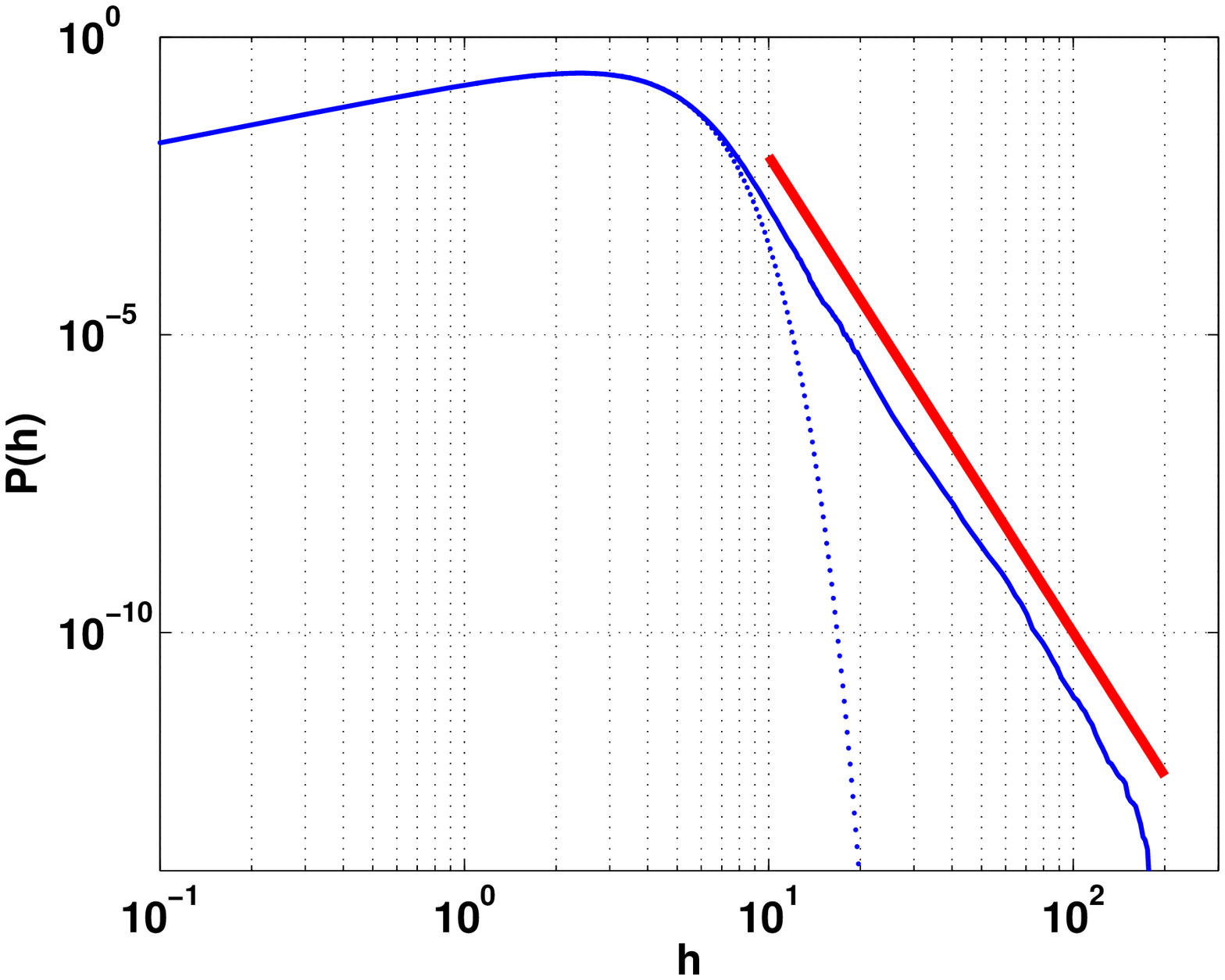} 
\caption{(Color online) PDFs for simulations with $a=1,c=1,b=10^4,\epsilon= 2\cdot 10^{-3}$. (a) PDF ${\cal P}_r(h_r)$ for  $h_r=\mathrm{Re}(\psi)$.
Solid blue line represents numerical result and dotted blue line is for Gaussian distribution $(2\pi)^{-1/2}h_0^{-1}e^{-h^2/(2h_0^2)}$ plotted for comparison.
Here the variance $h_0^2 = 5.85$ is obtained from the simulation.
(b) PDF ${\cal P}(h)$ of $h=|\psi|$ at linear scale. Solid blue line shows numerical result and dotted blue line is Gaussian distribution
$h_0^{-2} he^{-h^2/(2h_0^2)}$, where $h_0^2$ is the same variance as in (a).
(c) PDF ${\cal P}(h)$ of $h=|\psi|$ at log-log scale. Solid blue line shows numerical result and dotted blue line is the Gaussian distribution as in (b). Thick solid red line shows $h^{-8}$ law for comparison.}
\label{re}
\end{center}
\end{figure}

Figure
\ref{re}a
shows PDF ${\cal P}_r(h_r)$ for the
real part of the amplitude $\psi$ to have a value $h_r$. PDF for the imaginary part of $\psi$ has the same form. Solid line is the match to the Gaussian distribution (with the same variance as for ${\cal P}_r(h_r)$)
which is almost indistinguishable from
${\cal P}_r(h_r)$. PDF ${\cal P}_r(h_r)$ is
determined from simulations as
\begin{equation}\label{PDFrepsidef}
{\cal P}_r(h_r)=\frac{\int \delta (Re(\psi(x,t))-h_r)dx dt}
{\int dxdt}.
\end{equation}
Here the integrals are taken over all values of $x$ and all
values of $t$ after the turbulence has reached the statistically
steady state. We assume ergodicity of turbulence, i.e., that averaging over space and time is equivalent to the averaging over ensemble of initial conditions (or the stochastic realizations of random forcing for the
random forcing case).

In a similar way, ${\cal P}(h)$ for the amplitude $|\psi|$ to have a value $h$ is
determined from simulations as
\begin{equation}\label{PDFabspsidef}
{\cal P}(h)=\frac{\int \delta (|\psi(x,t)|-h)d{x} dt}
{\int d{x}dt}
\end{equation}
and, as shown in Figure
\ref{re}b, it is again very close to the Gaussian distribution.
Large fluctuations of $|\psi|$ are however, quite different from the Gaussian distribution and have power-like tails as shown in Figure \ref{re}c at log-log scale.
From comparison of Figures \ref{re}a, b and c we conclude that the fit to the Gaussian distribution
works very well for $|\psi|\lesssim 8$ which can be interpreted as the superposition of numerous almost
linear waves. For $|\psi|\gtrsim 10$ the PDF has a power law-like
dependence which can be roughly estimated as $\sim h^{-8}$ and which indicates  intermittency of the optical turbulence
\cite{ChungLushnikovVladimirovaAIP2009}.

\begin{figure}[ptb]
(a)
\includegraphics[width = 3.4 in]{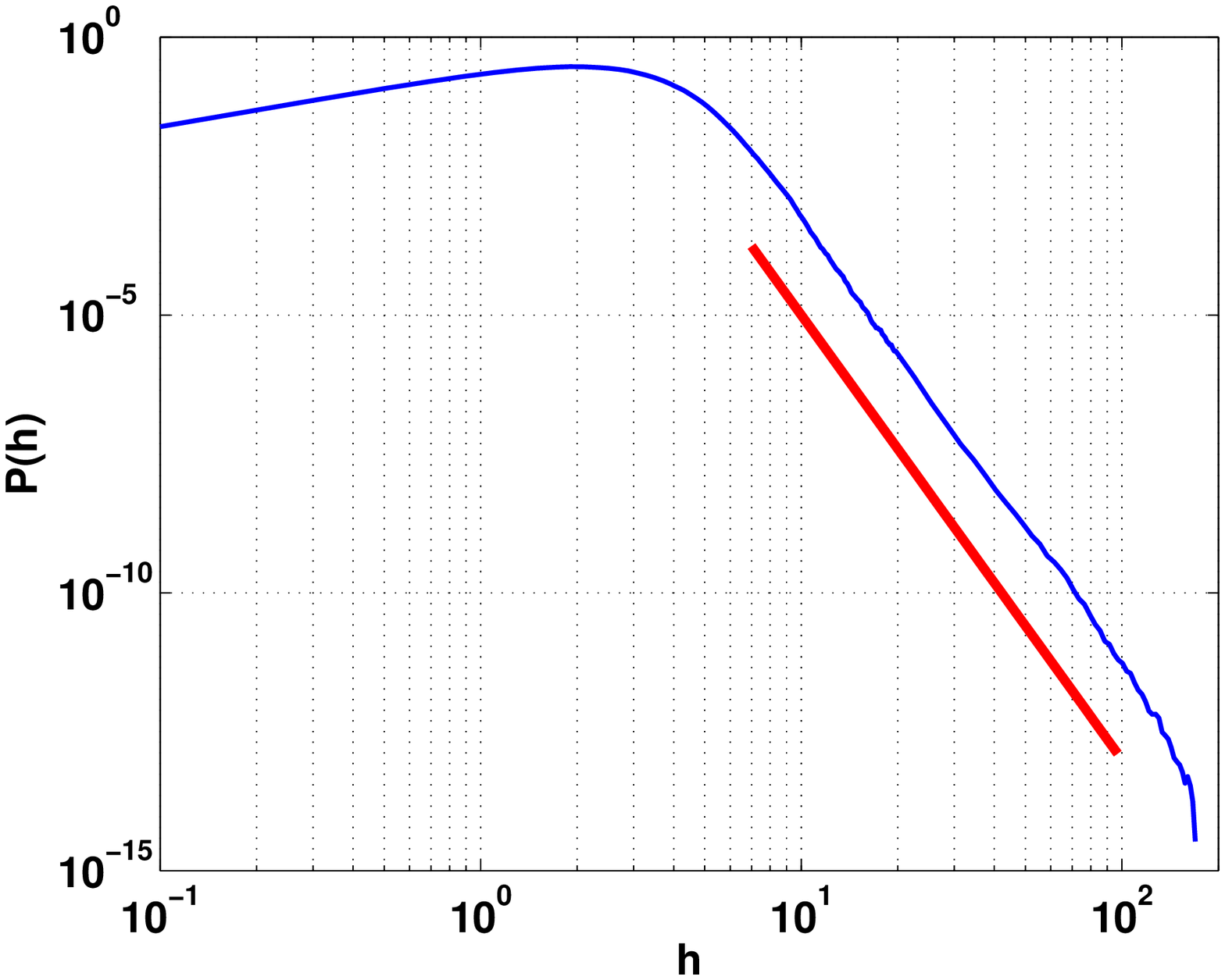} 
(b)
\includegraphics[width = 3.4 in]{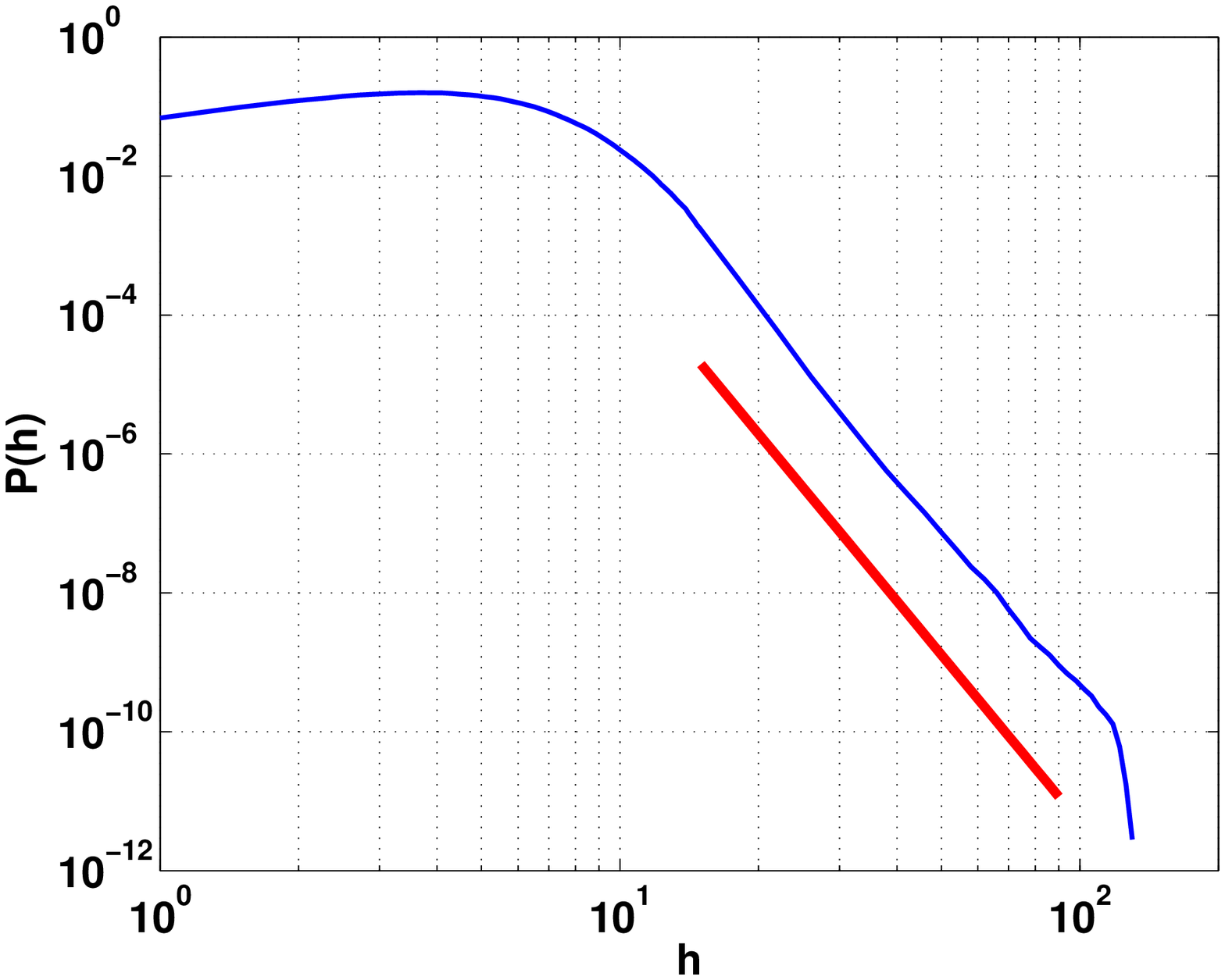} 
\caption{(Color online) (a) PDF ${\cal P}(h)$ of $h=|\psi|$ for ``$k$-limited" deterministic forcing (\ref{linearforcinggeneral}), (\ref{bkdefcutoff}) with $k_{cutoff} = 10\pi$.
Parameters are $a=c=1, b_0=10^4, \epsilon=2 \cdot 10^{-3}$. (b) PDF ${\cal P}(h)$ of $h=|\psi|$ for random forcing with parameters  $a=c=1,\epsilon=10^{-3},l_c=0.02,b_g=12500$.}
\label{force10m}
\end{figure}

Figure \ref{force10m}a shows the PDF ${\cal P}(h)$ for RQNLS with the ``$k$-limited" deterministic forcing  (\ref{linearforcinggeneral}),  (\ref{bkdefcutoff})
and Figure \ref{force10m}b shows the PDF ${\cal P}(h)$ for RQNLS
with the random forcing (\ref{stochasticforcing}). It is seen that both of these cases have the same type of power-like tails  as in Figure  \ref{re}c. Thus, the power-like tails are universal for RQNLS for $\epsilon\ll 1$
and independent of the type of forcing indicating the universal turbulent behavior.

\begin{figure}
(a)
\includegraphics[width = 3.4 in]{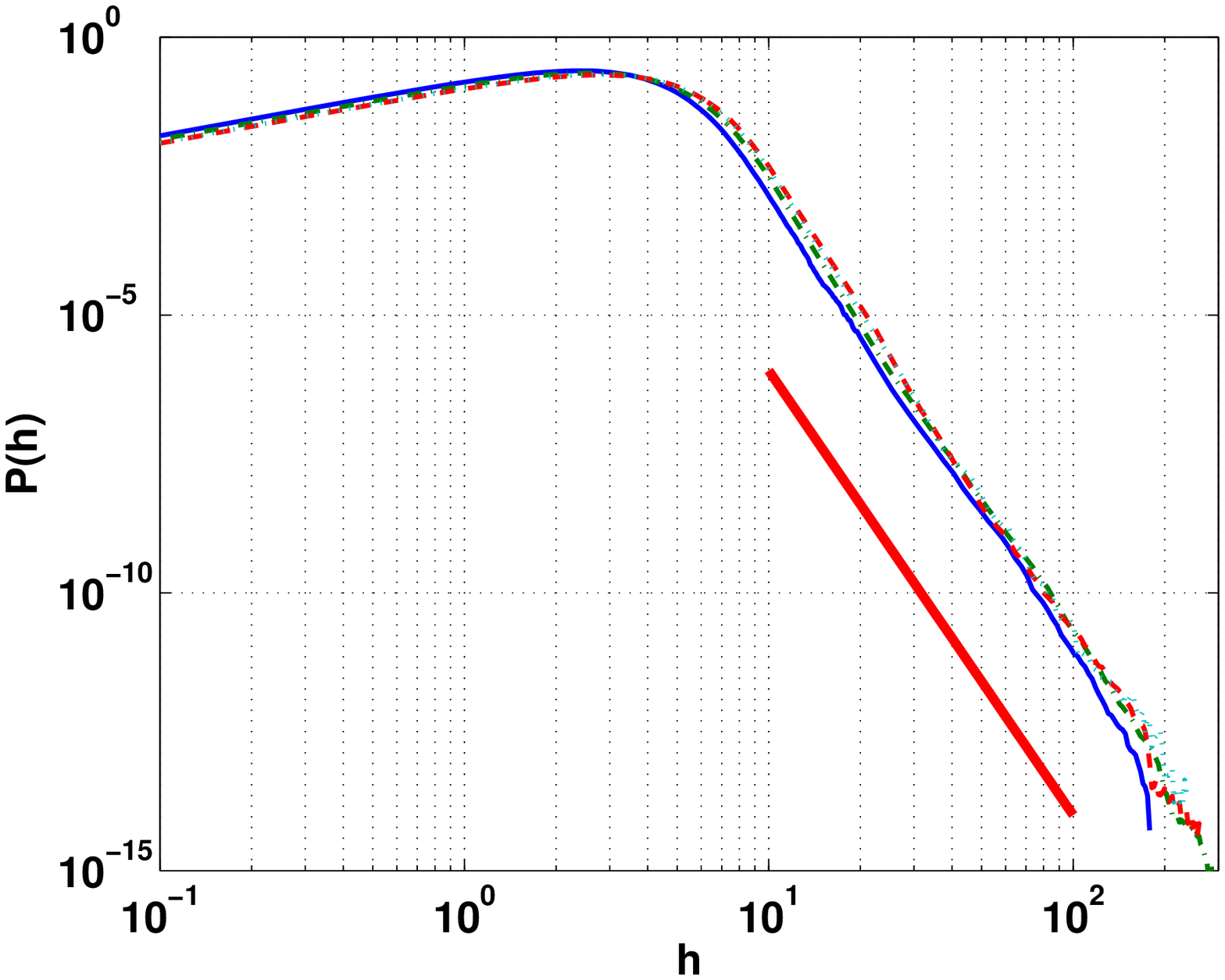} 
(b)
\includegraphics[width = 3.4 in]{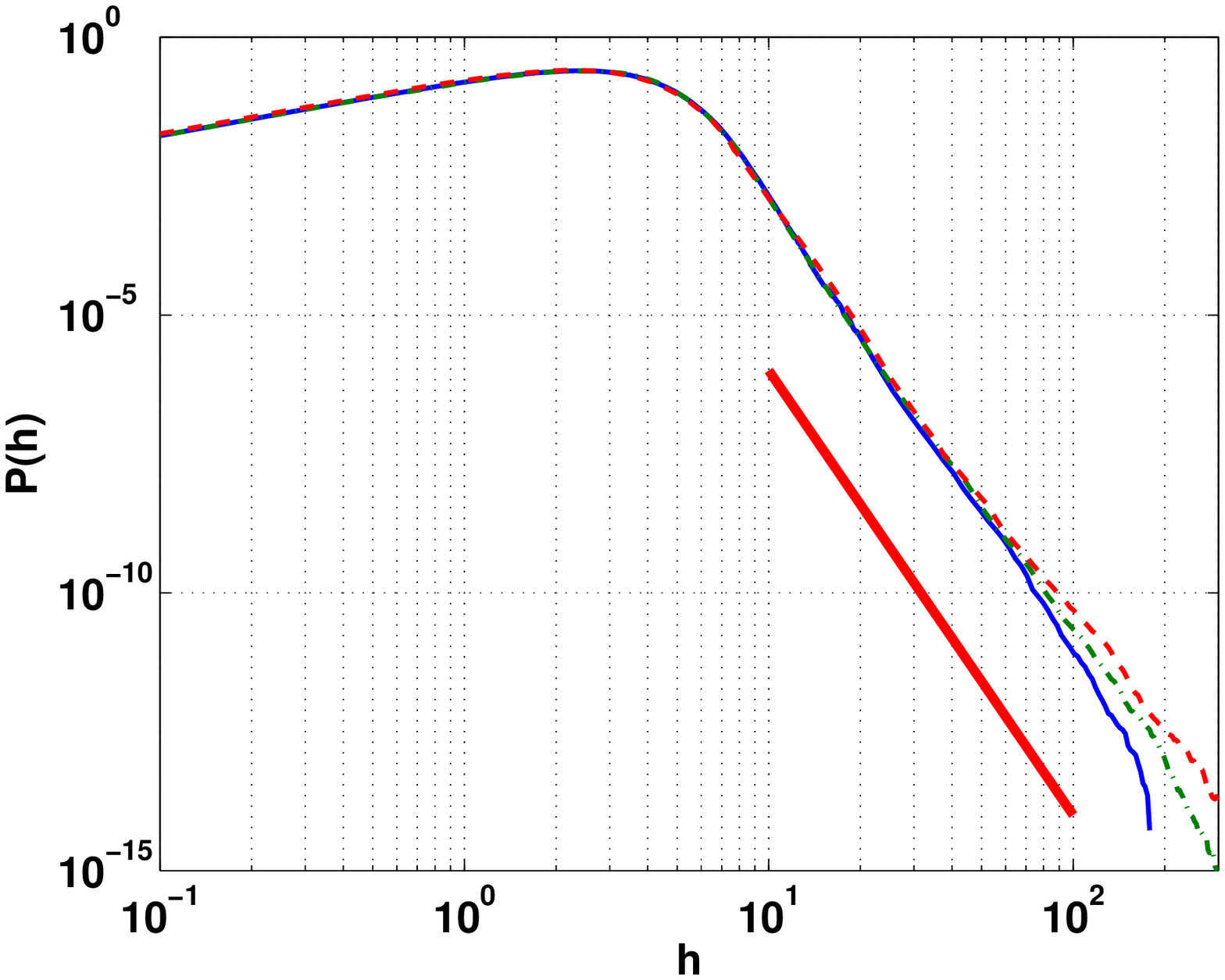} 
(c)
\includegraphics[width = 3.4 in]{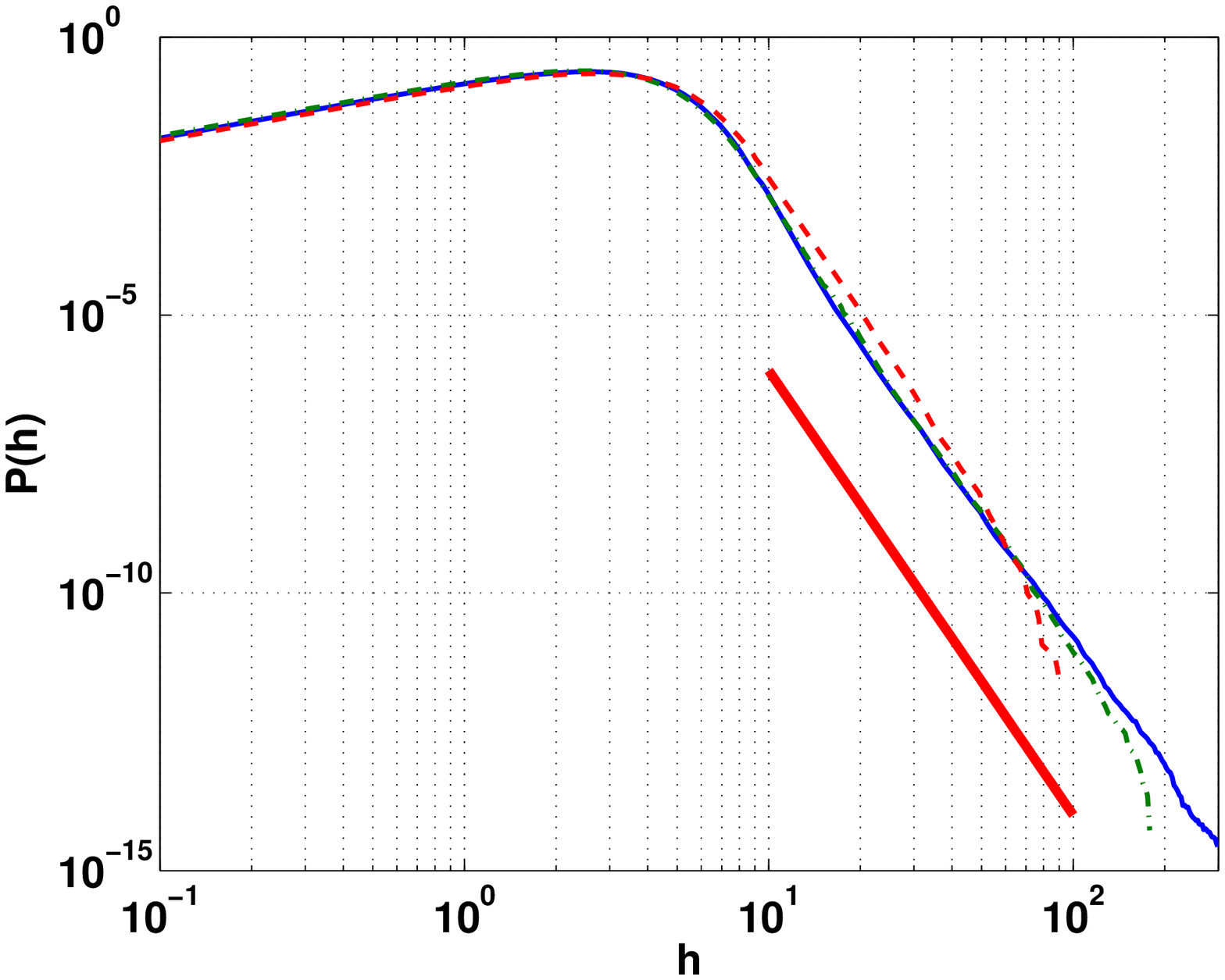} 
\caption{(Color online) PDF ${\cal P}(h)$ of $h=|\psi|$. (a) Parameters are $\epsilon=2 \cdot 10^{-3},c=1,b=10^4$, $a=0.05$ (dotted), $a= 0.1$ (dashed), $a=0.5$ (dashed-dotted), $a=1$ (solid).
(b) Parameters are $\epsilon = 2 \cdot 10^{-3},a=1,b=10^4$, $c= 0.1$ (dashed), $c=0.5$ (dashed-dotted), $c=1$ (solid).
(c) Parameters are $a=1,c=1,b=10^4,\epsilon= 10^{-3}$ (solid), $\epsilon=2\cdot 10^{-3}$ (dashed-dotted), $\epsilon=5 \cdot 10^{-3}$ (dashed). Thick solid line shows $h^{-8}$ power law in all Figures.}
\label{P:a}
\end{figure}

In order to characterize the relevance of linear and nonlinear dissipation in PDF ${\cal P}(h)$,  we consider the different parameter values for $a$ and $c$
while fixing $\epsilon$ and $b$.
Figure \ref{P:a}a shows   ${\cal P}(h)$ for
fixed nonlinear dissipation  $c=1 $ and
different values of the linear dissipation coefficient $a$. As $a$ decreases, the slope of PDF tail becomes steeper deviating from  $h^{-8}$ power law.
For $a<0.1$, we observe that the shape of  ${\cal P}(h)$ tail is very close to the case of $a=0.1$.
Figure \ref{P:a}b shows   ${\cal P}(h)$ for fixed linear dissipation $a=1$ and variable $c$.
 The PDF tail again deviates from $h^{-8}$ power law as  $c$ decreases. The change of tail is more significant at large amplitudes $h$.
 As $c$ becomes smaller, high amplitude collapses occur more often, e.g., for $c \sim 0.05$, we frequently observe collapses with amplitude $\geq 500$.
%
In contrast, if we change $\epsilon$ only, then the tails of  ${\cal P}(h)$ behave similarly for small $\epsilon \lesssim  3\cdot 10^{-3}$ except very large values of $h$ as shown in Figure \ref{P:a}c.

\begin{figure}
\includegraphics[width = 3.4 in]{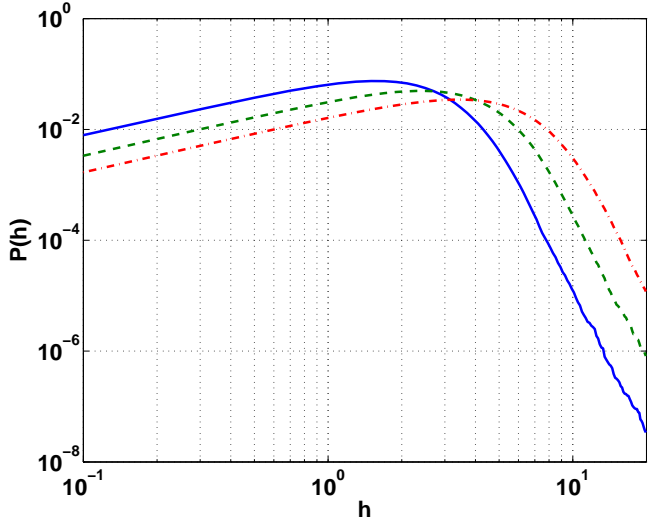} 
\caption{(Color online) Zoom-in of PDF ${\cal P}(h)$ on the range of small $h=|\psi|$. Parameters are $\epsilon=2\cdot 10^{-3}, a=c=1$, $b=2\cdot 10^3$ (solid), $b= 10^4$ (dashed), $b=5\cdot 10^4$ (dashed-dotted). }
\label{ave_R}
\end{figure}

We also study the sensitivity of ${\cal P}(h)$ to the change of amplification amplitude $b$.
Figure  \ref{ave_R} shows that a height of the maximum of  ${\cal P}(h)$ decreases with the increase of $b$ while the position of the maximum $h=h_0$ shifts to the right indicating the increase of the average amplitude $\langle |\psi|^2\rangle$ (see also Figure \ref{N_corr}). To stress this feature we focus in Figure  \ref{ave_R} on a smaller domain in $h$ and compare with previous figures, while
for larger $h$ the tails of ${\cal P}(h)$ behave similar for different $b$ up to the normalization constant.

\subsection{Power law-like tails of PDF}
\label{section:PDFtails}

We now show that the power-like tail of ${\cal P}(h)$ results from the
near-singular collapsing events.  This approach dates back to the idea of
describing strong turbulence in the Navier-Stokes equations through
singularities of the Euler equations \cite{FrischBook1995}.
Unfortunately, this hydrodynamic problem remains unsolved.  The forced
Burgers equation remains as the only example of an analytical description
of strong turbulence in which the tail of the PDF for negative
gradients follows a well established $(-7/2)$ power law \cite{EKhaninMazelSinaiPRL1997},
dominated by the dynamics of near-singular shocks. Another example of the analytical description of the intermittency
is a randomly advected  passive scalar which is the example of the turbulent transport described by  linear equations \cite{ChertkovFalkovichKolokolovLebedevPRE1995}.


As a first step we calculate the contribution to the PDF  from individual collapse events.
As shown in Figure \ref{coll_over}, the filament
amplitude $|\psi|_{max}$ rapidly decays after reaching $|\psi|_{maxmax}$ at $t=t_{max}$. Thus, we
neglect the contribution to ${\cal P}(h)$ from $t\gtrsim t_{max}$ in calculating the contribution of the individual filament to ${\cal P}(h)$.
We define the conditional probability ${\cal P}\left(h|h_{max}\right)$ for contribution to PDF from the collapse event with $|\psi|_{maxmax}\equiv
h_{max}$ and
use (\ref{selfsimilar}) ,(\ref{R1D}), (\ref{gammadef}) and (\ref{PDFabspsidef})
as follows
 \begin{eqnarray}
 \begin{split}
&{\cal P}\left(h|h_{max}\right)\propto \int\limits
^{t_{max}} dt \int d{x}\delta\left(h-\frac{1}{L(t)^{1/2}}R_0\left(\frac{x}{L(z)}\right)\right) \\
&\propto
\int d{
\rho}\, \int\limits^{L(t_{max})} \frac{dL \,
L^{2}}{\gamma}\delta\left(h-\frac{1}{L(t)^{1/2}}R_0(\rho)\right) \\
&\simeq\int
\frac{d{\rho}\,}{{ \langle \gamma \rangle}h^{7}}
{ [R_0(\rho)]^{6}}
\Theta\left(\frac{R_0(0)}{L(t_{max})^{1/2}}-h\right)
 \\
&= Const \  h^{-7} \ \Theta\left(h_{max}-h\right),
\label{P_L_rho}
\end{split}
\end{eqnarray}
where $h_{max}=R_0(0)/L(t_{max})^{1/2}=3^{1/4}/L(t_{max})^{1/2}$, and $\Theta(x)$ is the Heaviside
step function.  Here, we have changed the integration variable from $t$ to
$L$ and approximated $\gamma(t)$ under the integral by its average
value $\langle \gamma \rangle$ as $\gamma(t)\simeq \langle \gamma
\rangle\sim0.5$.  This approximation is valid for $t\lesssim
t_{max}$ outside the neighborhood of $t=t_{max}$ as seen in Figure \ref{gamma}.

As a second step we calculate ${\cal P}(h)$ by integration over all
values of $h_{max}$ using equation (\ref{P_L_rho}) as follows
\begin{eqnarray}\label{hpower}
{\cal P}(h)&=&\int  dh_{max}{\cal P}(h|h_{max}) {\cal P}_{max}(h_{max})
\nonumber
\\
&\simeq&  Const \ h^{-7}\int dh_{max} \Theta(h_{max}-h) {\cal P}_{max}(h_{max})\nonumber \\
&=& Const \ h^{-7} H_{max}(h),
\end{eqnarray}
where ${\cal P}_{max}(h_{max}) $ is the PDF for $h_{max}=|\psi|_{maxmax}$
and $H_{max}(h)\equiv \int^\infty_{h} {\cal P}_{max}(h_{max})
dh_{max}$ is the cumulative probability for $|\psi|_{maxmax}>h$.

\begin{figure}
(a) \includegraphics[width = 3.4 in]{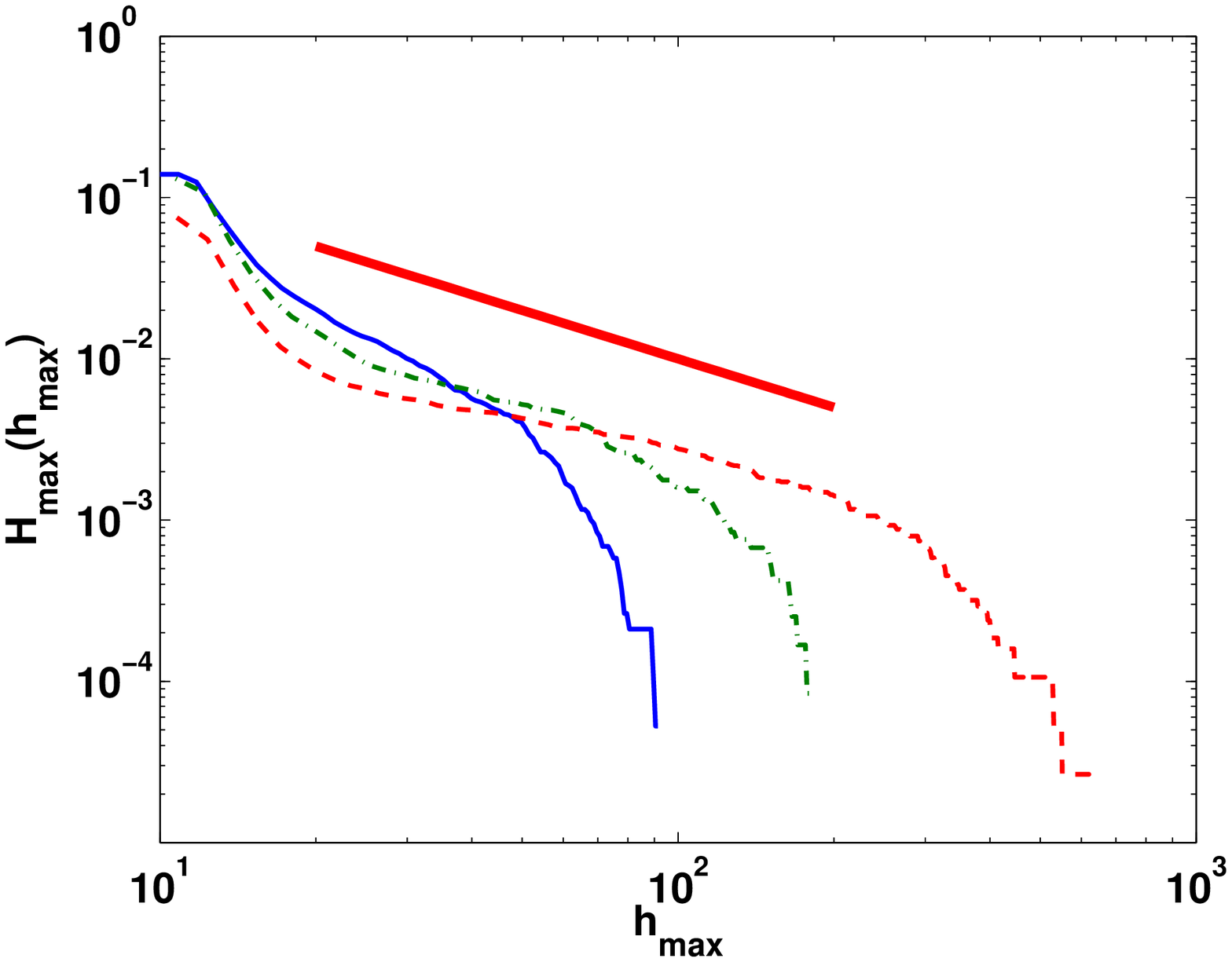} 
(b) \includegraphics[width = 3.4 in]{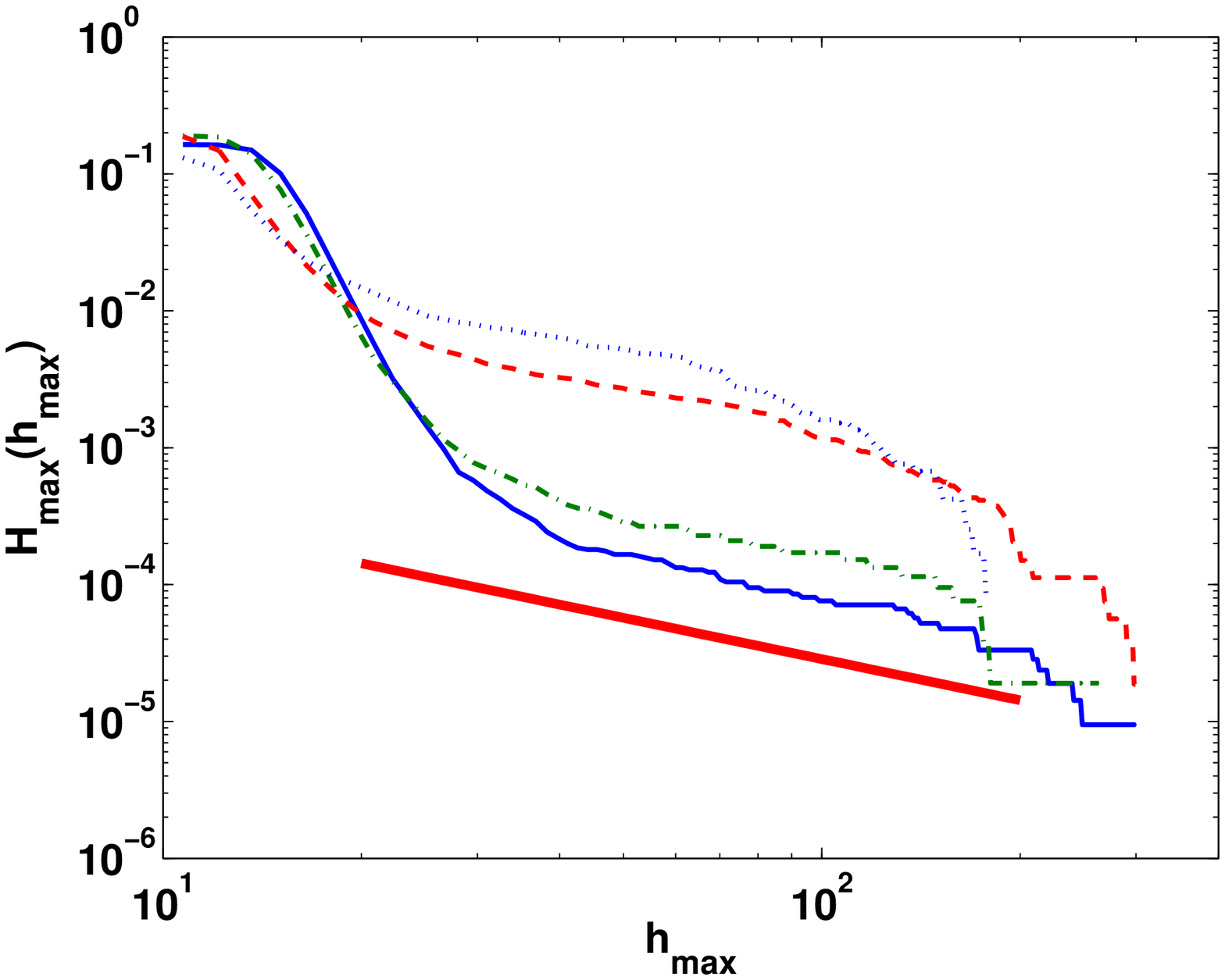} 
\caption{(Color online) Cumulative probability $H_{\max}(h_{max})$ for $h_{max}=|\psi|_{max}$ and deterministic forcing with $c=1,b=10^4$. (a) Parameters are $a=1,\epsilon=5\cdot 10^{-3}$ (solid), $\epsilon=2\cdot 10^{-3}$ (dashed-dotted), $\epsilon=10^{-3}$
 (dashed).
(b) Parameters are $\epsilon= 2 \cdot 10^{-3}$. $a=0.05$ (solid), $a=0.1$ (dashed-dotted), $a=0.5$ (dashed), $a=1$ (dotted).
Thick solid red line shows  $h_{max}^{-1}$ power law for comparison in both figures.}
\label{G:eps}
\end{figure}

Figures~\ref{G:eps}a and \ref{G:eps}b show $H_{max}(h_{max})$ for different values of parameters.
Each curve is calculated after the system reaches the statistical steady using more than $10^3$ collapse events with $h_{max}>10$. We verified that the increase of the number of collapse event
(i.e., increase of the total simulation time) does not change these curves in any significant way.
Figure~\ref{G:eps}a shows that power-like dependence $h_{max}^{-1}$ for $H_{max}(h_{max})$ if exist at all, quickly disappear with the increase of $\epsilon.$
Figure \ref{G:eps}b shows that  $H_{max}(h_{max})$ shows that power-like dependence $h_{max}^{-1}$ shifts to larger values $h_{max}$ with the decrease of $a.$
Generally, we see from  Figures~\ref{G:eps}a and \ref{G:eps}b that for a wide range of parameters,
including the decrease of $\epsilon$, the dependence of $H_{max}(h_{max})$ cannot be approximated  as $\propto (h_{max})^{-1}$.
The assumption of $H_{max}(h_{max})\propto (h_{max})^{-1}$ comes from (\ref{hpower}) and the rough estimate that ${\cal P}(h)\propto h^{-8}$ as in thick solid red line of Figure \ref{re}c.
We conclude that this conjecture, first made in \cite{IwasakiTohProgrTheorPhys1992}, appears to be incorrect and $h^{-8}$ is only a very crude approximation for
${\cal P}(h)$.

\begin{figure}
(a)
\includegraphics[width = 3.4 in]{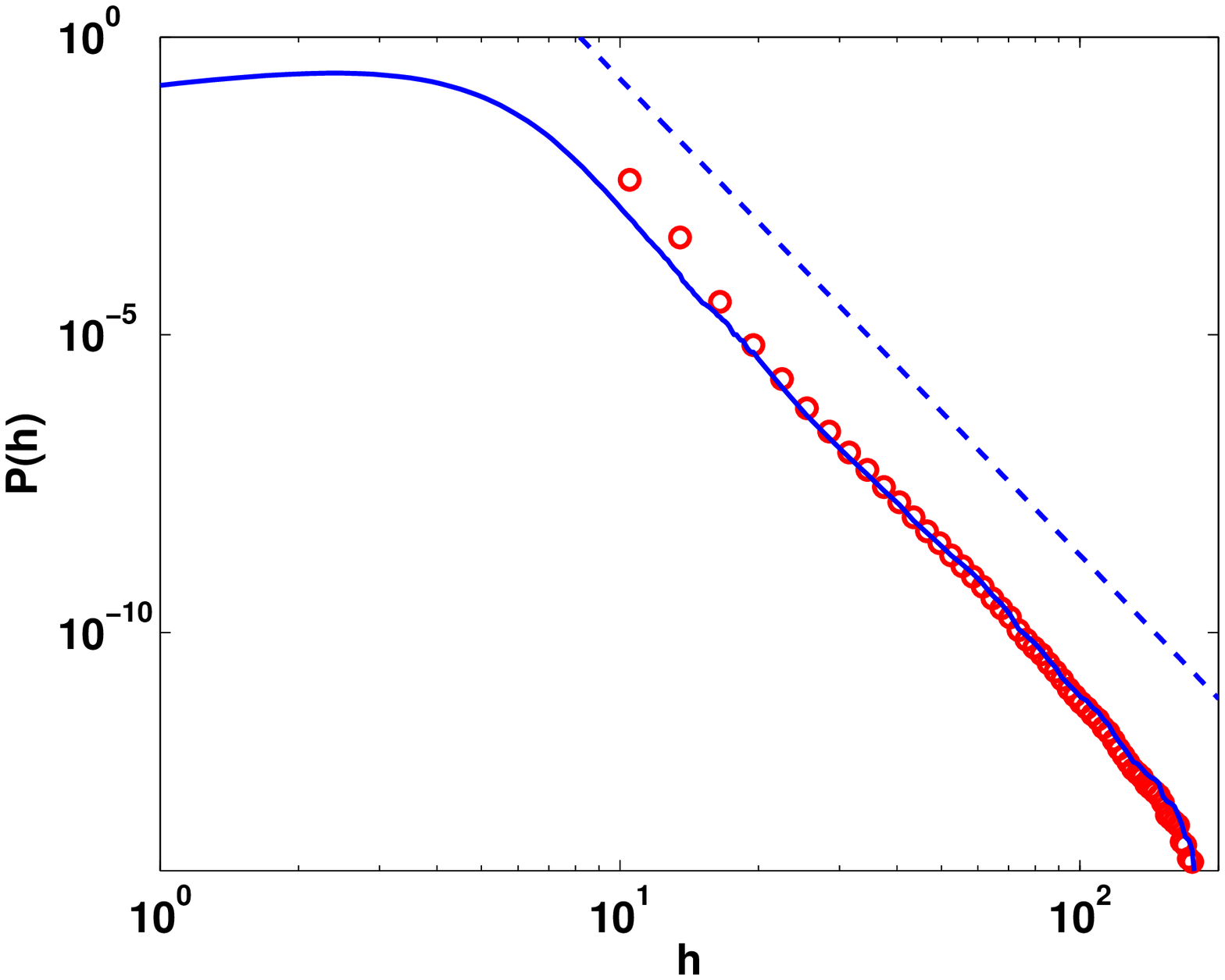} 
(b)
\includegraphics[width = 3.4 in]{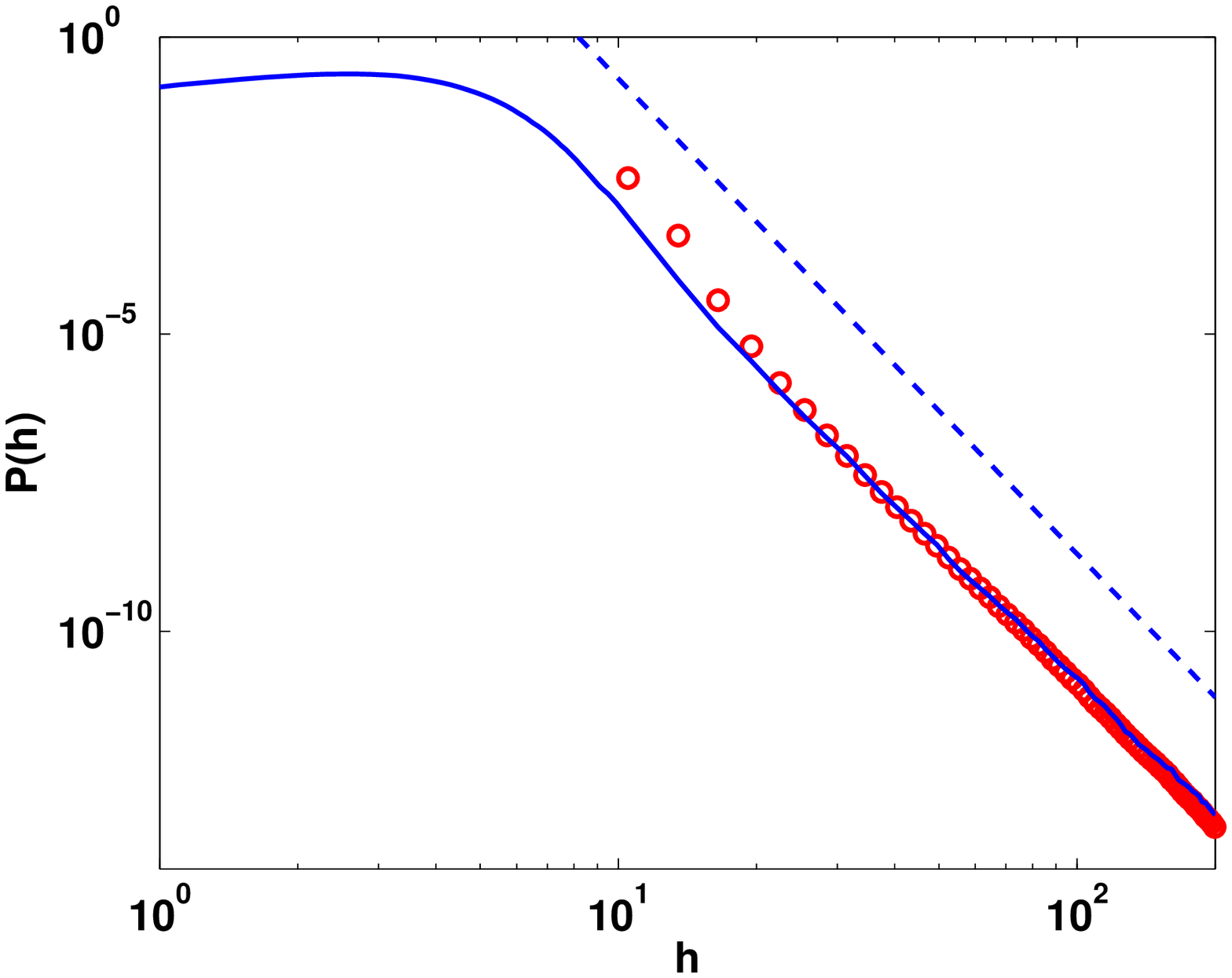} 
(c)
\includegraphics[width = 3.4 in]{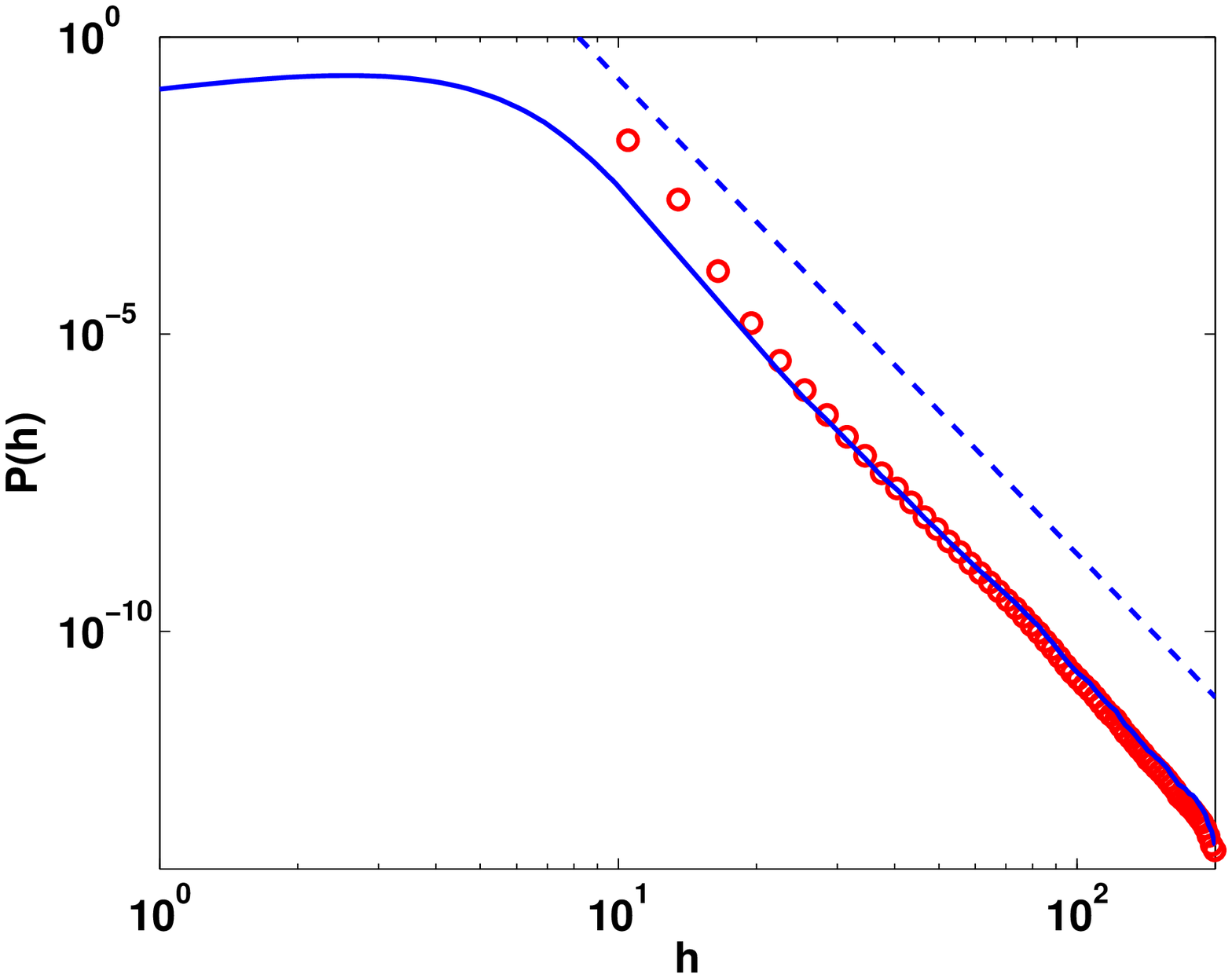} 
\caption{(Color online) Solid curve represents PDF ${\cal P}(h)$ for $h=|\psi|$, and dashed line shows $h^{-8}$ power law. Circles correspond to the rhs of equation~(\ref{hpower}).
(a) Parameters are $\epsilon= 2 \cdot 10^{-3},a=c=1,b=10^4$.
(b) Parameters are $\epsilon= 10^{-3},a=c=1,b=10^4$.
(c) Parameters are $\epsilon= 2 \cdot 10^{-3},a=0.5,c=1,b=10^4$.
}
\label{pdf_fit}
\end{figure}

We now verify that the equation  (\ref{hpower}) is correct one, which justifies the assumptions used in (\ref{P_L_rho}) and   (\ref{hpower}).
Figures \ref{pdf_fit}a, \ref{pdf_fit}b and \ref{pdf_fit}c compare ${\cal P}(h)$ from simulations (solid blue lines) with the prediction of
the equation~(\ref{hpower}) (red circles), where $H_{max}(h_{max})$ is obtained numerically and shown in Figure \ref{G:eps}.  Also Figure \ref{G:eps} shows  that for   $h\gtrsim 20$
the equation~(\ref{hpower})
appears to be much better fit of  ${\cal P}(h)$ compare with
 $h^{-8}$ power law.
The normalization constant in the equation~(\ref{hpower})
was chosen to fit  ${\cal P}(h)$ at large $h$.  The deviation of (\ref{hpower}) from ${\cal P}(h)$ for $h\lesssim 20$ is perhaps due to
 the decaying of the large amplitude collapse events into large amplitude waves (as e.g. in Figure \ref{rho}c). This explanation is consistent with the increase of that deviation
 for smaller values of $a$ as in Figures \ref{pdf_fit}c because the decrease of $a$ causes these large amplitudes almost linear waves to live longer before dissipating, thus giving a bigger contribution
 into ${\cal P}(h)$. We also conclude from Figures \ref{pdf_fit}a, \ref{pdf_fit}b and \ref{pdf_fit}c that because such deviation is insignificant for $h\gtrsim 20$, the contribution of large amplitude waves is not important in that range of $h$.  The  good agreement between
${\cal P}(h)$ and the equation~(\ref{hpower}) justifies the assumptions used in derivation
of the equation~(\ref{hpower})
as well as it shows
that the intermittency of optical turbulence of  RQNLS
(\ref{nlsregul}) is solely due to collapse dynamics.
We also conclude that while  $h^{-8}$ power law appears to be an intermediate fit at best, the equation  (\ref{hpower})  works for all values of parameters of RQNLS we tested.


\section{Numerical simulations of 1D RQNLS}

In this section we provide detailed description of the numerical methods used to produce the simulations described above. We conduct numerical simulation of RQNL (\ref{nlsregul}) by employing a version of
the fourth order split-step method \cite{YoshidaPhysLettA1990} (outlined below)
for deterministic forcing as well as the second order split-step method for random forcing. For efficient computation, we adaptively
change the spatial grid size $\Delta x$ during the time evolution in the spatial domain  $-0.5\leq x \leq 0.5$. Specifically, if the amplitudes of Fourier components of solution,
$|\psi_k|$, at high frequencies exceed  $10^{-10}\max \limits _k|\psi_k|$, we reduce $\Delta x$ by adding more Fourier modes to the system.
If the amplitudes of high frequency modes are below the criteria, we increase $\Delta x$ by removing some of the existing Fourier modes.
The numerical time step, $\Delta t$, is also updated as
$\Delta x$ changes, which follows the relation $\Delta t k_{max}^2= q_0\pi$. Here $k_{max}=\pi/\Delta x$ is the maximum wavenumber determined by the discretization
  and we  choose the constant factor $q_0$ small enough to avoid  numerical instability.
The numerical instability occurs if the change of phase $\Delta \phi=\Delta t k_{max}^2$
of the highest Fourier harmonics $k_{max}$ from the linear term of RQNLS at one discrete time step $\Delta t$ is  above $\pi/2:$ $\Delta \phi\ge \pi/2$.
In that case a  mixing of the Fourier harmonics in the quintic nonlinear term in RQNLS can produce artificial (discretization-caused) resonance from the condition $4\Delta \phi=2\pi$.
It implies that to avoid the instability we have to choose $q_0<1/2$.
In simulation we typically choose $q_0=0.2$ which allows to avoid instability as well as  insures high accuracy in time stepping. The initial condition of our
 numerical simulation is a random field with small amplitude whose maximum value is about $1.6$.


\subsection{Fourth order Split-Step scheme}
We write equation (\ref{nlsregul}) formally in the form of

\begin{equation}\label{operatorform}
\psi_t = (\hat L+\hat N)\psi,
\end{equation}
where $\hat L$ is  the linear and $\hat N$ is the nonlinear operators, defined by
\begin{eqnarray}
\hat L \psi &=& i(1-ia\epsilon)\nabla^2\psi + \epsilon\phi, \label{lin_operator}\\
\hat N \psi&=& i(1+ic\epsilon)|\psi|^{4}\psi. \label{nonlin_operator}
\end{eqnarray}

Separately both  (\ref{lin_operator}) and  (\ref{nonlin_operator}) can be solved very efficiently. Here, we approximate the exact solution $\psi(t_0+\Delta t) = e^{\Delta t(\hat N+\hat L)}\psi(t_0)$ of (\ref{operatorform})
over one segment from $t_0$ to $t_0+\Delta t$ by the following expression
\begin{eqnarray}\label{splitstep4thorder}
&\psi(t_0+\Delta t) = e^{c_4 \Delta t\hat N} e^{d_3 \Delta t\hat L} e^{c_3 \Delta t\hat N} e^{d_2 \Delta t\hat L}e^{c_2 \Delta t\hat N} \nonumber\\
&\times e^{d_1 \Delta t\hat L}e^{c_1 \Delta t\hat N}\psi(t_0),
\end{eqnarray}
where
\begin{eqnarray}\label{cddef}
     &c_1=\frac{1}{2(2-2^{1/3})}, \quad
     c_2=\frac{1-2^{1/3}}{2(2-2^{1/3})}, \nonumber\\
     &c_3=c_2, \quad
     c_4=c_1,  \nonumber \\
     &d_1=\frac{1}{2-2^{1/3}},  \quad
     d_2=\frac{-2^{1/3}}{2-2^{1/3}},  \quad d_3=d_1.
\end{eqnarray}
The split-step expression  (\ref{splitstep4thorder}), (\ref{cddef}) is of the fourth order accurate and it is a straightforward generalization of the
forth order symplectic integration of  \cite{YoshidaPhysLettA1990} to non-Hamiltonian systems.

The linear part $e^{\Delta t\hat L}$ of the operator splitting  can be efficiently calculated by using the Fast Fourier Transform  algorithm (FFT) for the deterministic forcing (\ref{linearforcinggeneral}).
For the stochastic forcing (\ref{stochasticforcing}) the equation (\ref{lin_operator}) has a form of the inhomogeneous linear differential equation. Homogeneous part of that equation we solve again using
FFT while the contribution of the inhomogeneous term is obtained by the numerical integration over $t$ for each $x$. We use the  trapezoidal rule for the integration which makes the scheme the
 second order in the random noise case. Therefore, instead of the fourth-order split-step algorithm (\ref{splitstep4thorder}), we use the standard second order split-step algorithm
 $\psi(t_0+\Delta t) = e^{(1/2)\Delta t\hat L} e^{\Delta t\hat N} e^{(1/2)\Delta t\hat L}\psi(t_0),$ in the random noise case.

The nonlinear part $e^{\Delta t\hat N}$ of the operator splitting can be solved exactly as follows. Denote the solution of the nonlinear part of RQNLS as  $\psi^{\hat N}$.
 It means that $\psi^{\hat N}(t_0+\Delta t)= e^{\Delta t\hat N}\psi^{\hat N}(t_0)$. According to RQNLS,   $\psi^{\hat N}$ satisfies the following equation:
\begin{equation}
\partial_t \psi^{\hat N} = (-c\epsilon+i)|\psi^{\hat N}|^{4} \psi^{\hat N}. \label{nl}
\end{equation}
 It implies that
 \begin{equation}
\frac{d}{dt}|\psi^{\hat N}|^2 = -2c\epsilon|\psi^{\hat N}|^6,
\end{equation}
and hence, we find
\begin{eqnarray}
&|\psi^{\hat N}(t)|^2 = 0,\;\;\;\text{if}\;\; |\psi^{\hat N}(t_0)|^2 = 0, \label{n1}\\
&|\psi^{\hat N}(t)|^2 = [{4c\epsilon (t-t_0) + |\psi^{\hat N}(t_0)|^{-4}}]^{-\frac{1}{2}},\text{otherwise}. \nonumber\\
&\label{n2}
\end{eqnarray}
Using equations (\ref{n1}),(\ref{n2}), we obtain the explicit solution of equation (\ref{nl}),
\begin{eqnarray}
\psi^{\hat N}(t_0+\Delta t)& =& \exp\left[\frac{-c\epsilon+i}{4c\epsilon}\ln [4c\epsilon \Delta t|\psi^{\hat N}(t_0)|^4 + 1]\right] \nonumber\\
&\times&\psi^{\hat N}(t_0). \label{nlsol}
\end{eqnarray}

In the case of random forcing (\ref{stochasticforcing}), at each time step we independently generate the random variable $\xi(x)$ by the Ornstein-Uhlenbeck process
\cite{Gardiner2004} as a function of $x$ with zero mean and correlation
length $l_c= 0.02$
that satisfies a stochastic differential equation, $\frac{d\xi}{dx} = -l_c^{-1}\xi(x) + \sigma (\Delta t)^{-1/2} \frac{dW(x)}{dx}$, where $\sigma >0$ and $W(x)$ denotes the Wiener process. This Ornstein-Uhlenbeck process
yields an exponential
correlation function (\ref{chidef}) with $b_g = \sigma^2 l_c/2$. Here the factor $(\Delta t)^{-1/2}$ multiplying $d W/dx$  ensures  $\delta$-correlation of $\xi$ in time for $\Delta t\to 0.$

\section{Conclusion}

In this paper we studied the strong turbulence in  1D RQNLS (\ref{nlsregul}).
In the statistical  steady-state (the state of developed turbulence) the dynamical balance is achieved  between forcing which pumps the number of particles $N$ into the system
 and both linear and nonlinear dissipation.
RQNLS has multiple collapse events randomly distributed in space and time. We found that in the state of developed turbulence the spatial  and temporal correlation functions have universal forms
independent of the particular values of parameters of RQNLS as well as independent of the type of forcing. In particular we considered the deterministic forcing with variable $k-$dependence as well as random forcing.
We found that the relation between the correlation length  and the average amplitude $p_0$ has a universal form  (\ref{p02xcorr}) determined by the
modulational instability scale.

PDF ${\cal P}(h)$ of amplitude fluctuations  is well approximated by the Gaussian distribution for $h\lesssim 3p_0$.
In contrast, ${\cal P}(h)$  for $h \gtrsim  4p_0$  has  the strongly non-Gaussian tail  with power-like behavior characterizing intermittency of strong collapse-dominated turbulence.
This tail is determined by the equation ~(\ref{hpower}) which includes the contribution $\propto h^{-7}$
 from  the universal spatio-temporal form of the collapse events as well as the contribution from the cumulative probability  $H_{max}(h)$, the probability of the maximum amplitude of collapse event that exceeds $h.$
 We show that $H_{max}(h)$ is not universal and depends on the parameters of RQNLS. For some range of parameters $H_{max}(h)$ can be roughly estimated as $\propto h^{-1}$ but it appears to be an
 intermediate asymptotic at best

An important problem to be studied in future work is to determine the analytical form of $H_{max}(h)$ from the parameters of RQNLS. This is a challenging problem and will require calculation of the optimal fluctuations of the background which seeds new collapses. Moreover, it will be necessary to find the relation
between the size of such optimal fluctuation and the maximum collapse event amplitude $|\psi|_{maxmax}$.

\section*{Acknowledgments}

Support of Y.C. was provided by NSF grant DMS
 0807131. Work of P.L. was supported by NSF grant DMS
 0807131 and DOE Grant 1004118.

\bibliography{biblionls,lushnikov}

\end{document}